# The Kepler and TESS Asteroseismic Consortia

## A Historical Overview

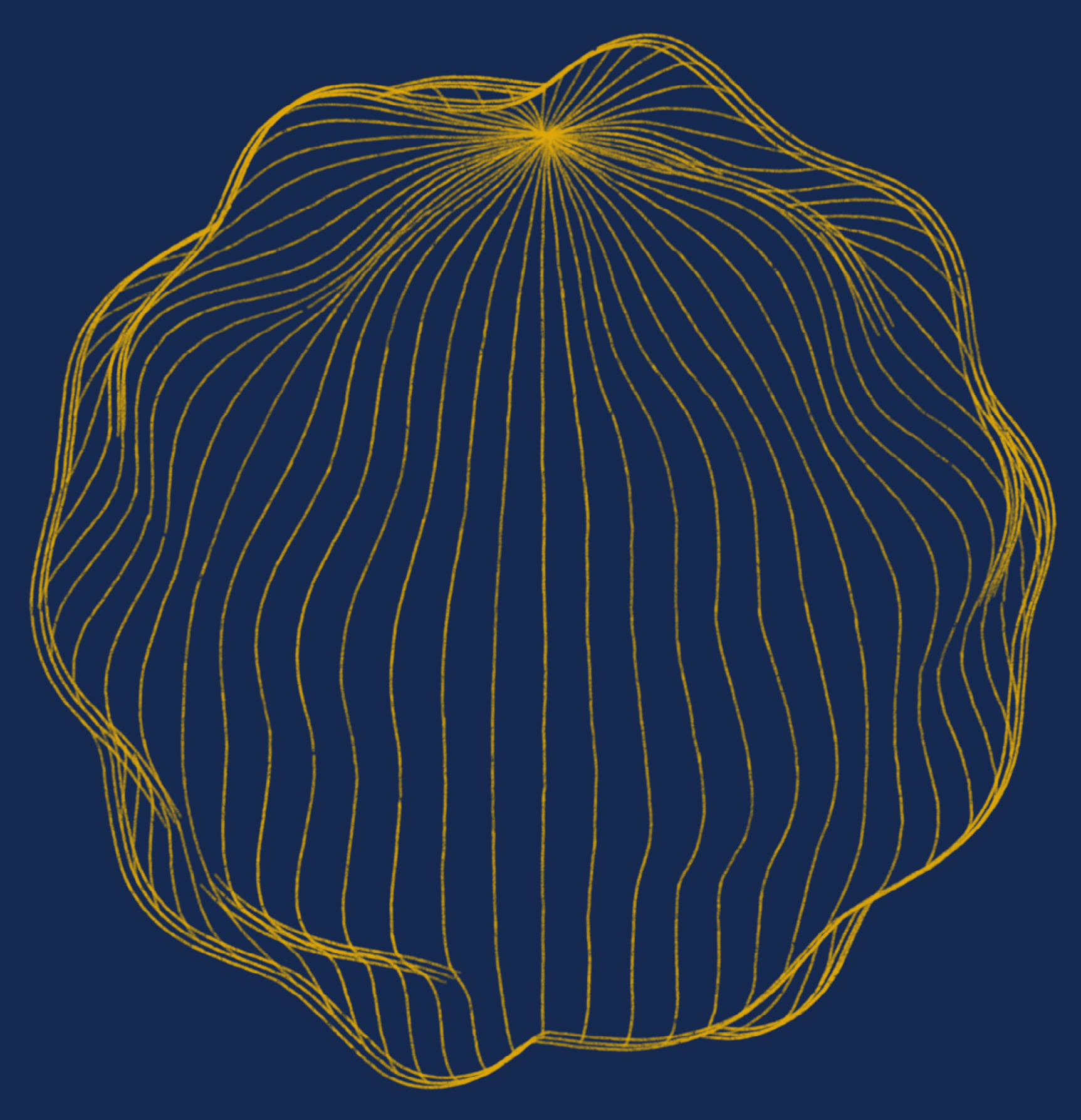

EDITED BY

*Mikkel N. Lund & Tiago L. Campante*

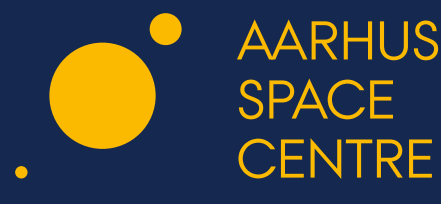

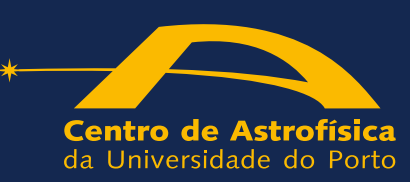

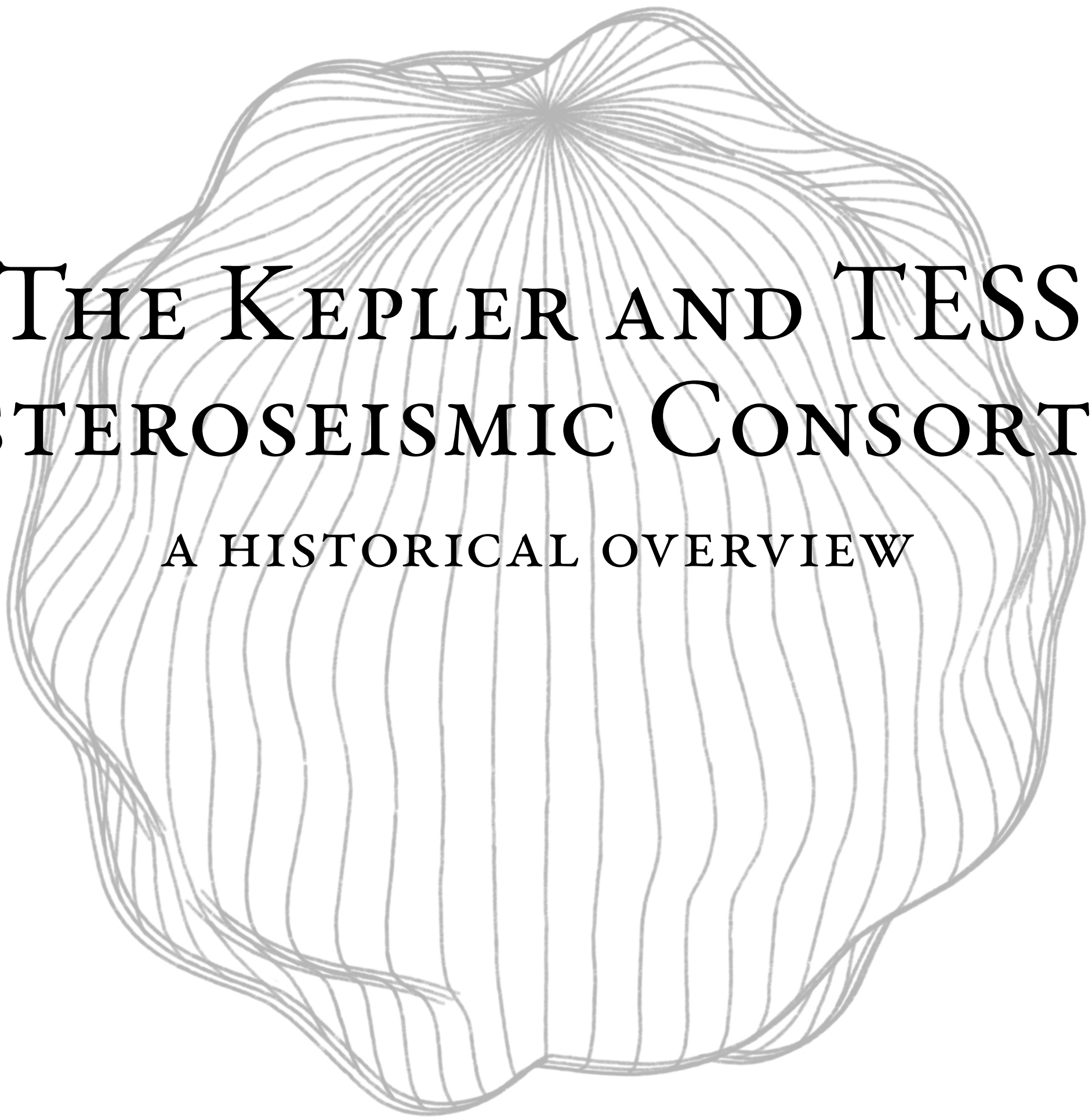

# The Kepler and TESS Asteroseismic Consortia

## A historical overview

edited by

*Mikkel N. Lund & Tiago L. Campante*

Editors:
Mikkel N. Lund
SpaCe – Aarhus Space Centre, Department of Physics and Astronomy
Aarhus University
Denmark

Tiago L. Campante
Centro de Astrofísica da Universidade do Porto
Instituto de Astrofísica e Ciências do Espaço
Portugal

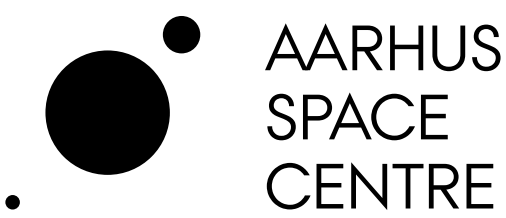

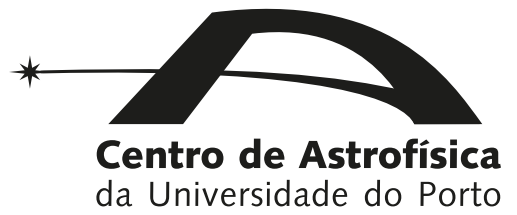

# Contents

# Preface

What is this about? In short: it is about people. About an idea that grew into a consortium, that grew into a community, that grew into a way of doing asteroseismology in the space era. The *Kepler* Asteroseismic Science Consortium (KASC) and later the TESS Asteroseismic Science Consortium (TASC) are not just organizational structures to access data. They became collaborative ecosystems in which students, postdocs, and senior scientists learned to work together across continents, time zones, and occasionally differing opinions about filters.

The idea for this booklet did not emerge from a formal meeting or a funding call. It emerged—appropriately enough for an international collaboration—over a beer. More precisely, over a beer (truth be told, more than one) at Beer Lab University in Honolulu, during the TASC7/KASC14 Workshop in 2023. Google Maps now informs us that this particular Beer Lab location is "permanently closed", which seemed at first like an ominous metaphor for fading inspiration. Fortunately, we have since learned that Beer Lab simply moved across the street, proving that good ideas, like good consortia, tend to reinvent themselves rather than disappear entirely. Somewhere between discussions that night and many others over the years, we found ourselves revisiting stories we had long heard from our former supervisors and more senior colleagues; stories of how KASC was negotiated, how the first working groups were formed, how data were filtered (a saga in its own right!), and how trust was built.

This booklet is therefore an attempt to capture some of that collective memory, while we still have access to the primary sources. It is also deliberately timed to celebrate the return of the KASC/TASC Workshop Series to Aarhus in 2026, the place where the consortia's original framework was conceived and set in motion. There is something fitting about reflecting on the journey of KASC and TASC as the community gathers again where it all began.

We are deeply grateful to everyone who helped shape the consortia into the collaborative and welcoming environment they have become. The scientific achievements of KASC and TASC are well known, but the spirit of collegiality and openness that underpins them is perhaps their most enduring legacy. In particular, we thank those who directly contributed to this historical overview by sharing their time, memories, and perspectives. Finally, thank you to all members, past and present, who made (and continue to make) KASC and TASC not just productive scientific collaborations, but communities in the fullest sense of the word. Unless explicitly attributed to a named contributor, all text in this volume reflects the views and recollections of the editors.

Mikkel Lund & Tiago Campante
Aarhus, February 2026

# I
# KASC and TASC

The *Kepler* Asteroseismic Science Consortium (KASC) and the TESS Asteroseismic Science Consortium (TASC) are large, international scientific collaborations established to coordinate the asteroseismic exploitation of NASA's *Kepler* and TESS space missions. Since their inception, these consortia have structured and fostered a vibrant community dedicated to advancing stellar astrophysics in particular, as well as exoplanet studies and Galactic archaeology (among other fields), through high-precision space photometry.

## The KASC

KASC was formed in 2007 following an agreement between the asteroseismology group in Aarhus, Denmark (specifically, Jørgen Christensen-Dalsgaard and Hans Kjeldsen) and the *Kepler* PI Bill Borucki, with significant influence from Ron Gilliland and Tim Brown. The mission's rationale for entering such an agreement was, first and foremost, that asteroseismic analysis could provide essential insights into the host stars of the mission's targeted exoplanets. In exchange for this contribution, access to the data would be granted, enabling the pursuit of asteroseismology with space-based photometry—something that was beginning in earnest at the same time with the French-led CNES/ESA CoRoT mission. For details and personal views on the events leading to the formation of KASC, please see the testimonials in § IV.

The asteroseismic work on *Kepler* data was organised under the "Kepler Asteroseismic Investigation" (KAI; see § D). The international community contributing to the KAI was organised within the KASC into a number (initially 14) of star-type-specific working groups (WGs), each with several subgroups. Each WG had an assigned chair, responsible for organising the asteroseismic analysis and providing inputs to the KAI on target selection. The KAI was governed by a small (4–5 person) steering committee, including members of the *Kepler* Science Team, responsible for overseeing the asteroseismic analysis and communicat-ing with the *Kepler* PI, Science Council, and Science Team. Below this top-level part of the organisation, the larger (initially 12 people) KASC Steering Committee, including some of the WG chairs, oversaw the day-to-day work of KASC and, importantly, also organised the new (and still ongoing) workshop series of the community (§III). All data sharing, paper reviews, and communication between WGs were facilitated via the *Kepler* Asteroseismic Science Operations Centre (KASOC; see §II).

While membership, and with it access to data, of the KASC has always been free for all, it did initially come with the requirement of a signed Non-Disclosure Agreement (NDA; see § C), to ensure that any accidentally identified exoplanets would not be published outside the *Kepler* team, and an acceptance of the KASC publication policies which included an internal review phase via KASOC. After the change in NASA's data release policy a few years into the mission, the NDA was dropped for new members, and the application of the infamous "transit removal filter" was discontinued (see the testimonials in § IV).

Since the early days of KASC, the number of WGs eventually reduced to 7—this structure continued to the end of the *Kepler* mission in 2018 (including the K2 mission), and stayed in effect for years after to organise the continued exploitation of *Kepler*/K2 data, until smoothly transitioning into the TASC with the launch of TESS.

### The TASC

With the preparations for the launch of the NASA/MIT Transiting Exoplanet Survey Satellite (TESS) well underway (eventually launched in 2018), an agreement was made between the TESS PI, George Ricker, and the KASC/KAI steering committees to continue the successful collaboration established with *Kepler*. This led to the formation of the TESS Asteroseismic Science Consortium (TASC) in 2015, closely following the final structure of KASC. Similarly, with TESS, the TESS Asteroseismic Science Operations Centre (TASOC) was created to serve as the database and communications platform of TASC. Under the TESS mission's fully open-data policy, membership in TASC required no NDA, only acceptance of the consortium's policies and publication strategy.

The working group structure that KASC had converged to was largely continued in TASC, but with the addition of a WG focusing on exoplanet hosts, a split of the main-sequence 'classical' pulsators into AF- and OB-types, and the removal of the WG on Mira and Semi-regular variables. Additionally, so-called "Coordinated Activities" (CAs) on 'TASC Data Products', 'TASC-GI Coordination', and 'Ground-based coordination' were introduced, specifically to address the lower level of processed data products to be delivered by the mission.

The management of TASC includes, at its top level, a Board—comprising Jørgen Christensen-Dalsgaard (TASC PI), Bill Chaplin, Steve Kawaler, George Ricker (TESS PI),

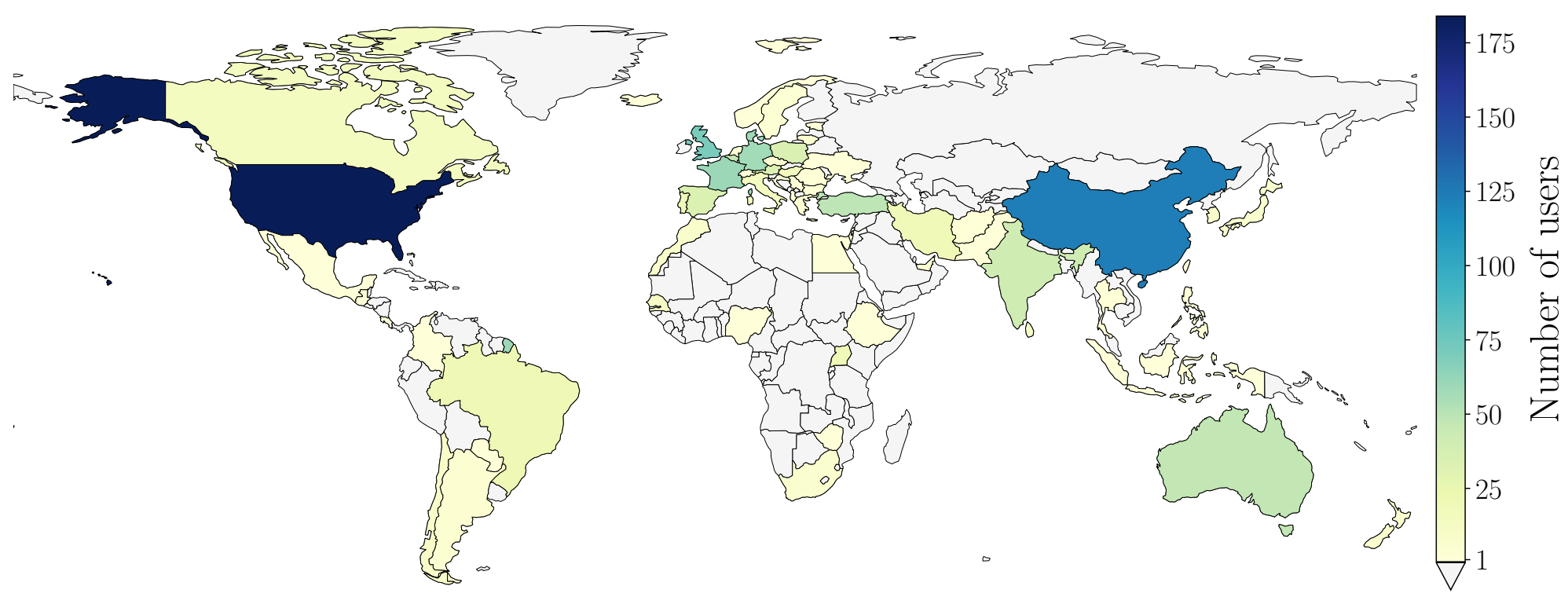

**Figure 1**: Geographical distribution of TASOC users by country (11 June 2026).

Roland Vanderspek (TESS Team), Sarbani Basu (selected by the US asteroseismic community), and Hans Kjeldsen (TASC Project Scientist)—with a focus on policies, WG structure, and ensuring collaboration. The TASC Steering Committee (including the Board, all WG/CA chairs and sub-chairs, and additional members selected for specific tasks) is in place to oversee the TASC asteroseismic analyses, ensure an optimal target selection, and, importantly, ensure the continuation of the KASC/TASC workshops.

Over the years, the membership of the consortia has steadily increased. Based on the number of users signed up on TASOC (§ II), the consortia include just over 1,000 individuals at the time of writing. The geographical distribution of these members is provided in Figure 1, which shows a broad representation across all (populated) continents. While many of the current TASOC users are likely students who do not continue in the field, the near-constant addition of $\sim$100 new members per year shows that the field is growing and flourishing—a trend we can only expect to continue in the future with a range of new missions and facilities on the horizon.

## Next steps...

At the time of writing, TESS continues to provide high-quality observations in its third extended mission (hopefully with many more to come), at observing cadences that have gradually been reduced thanks to the TESS team's impressive ability to continue improving data compression and transmission. To some extent, this has rendered many of the formal requirements for the WGs, such as actively providing input for target selection, moot. So, while there is still plenty of work to do on both *Kepler*/K2 and TESS data, and plenty of work is being done(!), the coordination of work within the TASC WGs has (with a few exceptions) gradually decreased.

This change in activity via TASC WGs is by no means a negative thing, but simply reflects a change in conditions on data availability. It also indicates that the consortium in the (near?) future might need a restructuring, and with many new missions suitable for asteroseismology—like PLATO, *Roman*, Earth 2.0, and hopefully HAYDN—on the horizon, a clean focus on asteroseismology with a more mission-agnostic approach might be prudent. The yearly KASC/TASC conferences remain the main event for asteroseismologists (and the likes) across the world to meet up, and in many cases reconnect with long-term colleagues and friends (as we did when coming up with the idea for this book), and are an excellent entry point for young researchers and students interested in the field. There is therefore no doubt that a consortium like TASC will remain relevant in the future.

# II
# The KASOC and TASOC databases

The *Kepler* and TESS Asteroseismic Science Operation Centres (KASOC and TASOC) are the data management and collaboration platforms made to support the work of the KASC and TASC. The development of these essential platforms for the asteroseismic community is the work of Rasmus Handberg. Having just started his PhD studies on the topic of 'Asteroseismology of Solar-like Stars' (Aarhus University, Denmark), he was tasked with developing a platform for data sharing, communication, publication review, and related activities for the newly formed KASC. While those who worked closely with Rasmus were well aware of his high level of technical and scientific expertise, it is fair to say that few could have anticipated the quality of the KASOC (and later TASOC) platform he built from scratch in record time, and kept maintaining, all the while continuing his scientific work on solar-like oscillators. Following his PhD and a postdoc period at the University of Birmingham, where he kept maintaining the databases, Rasmus eventually became the Data and Systems

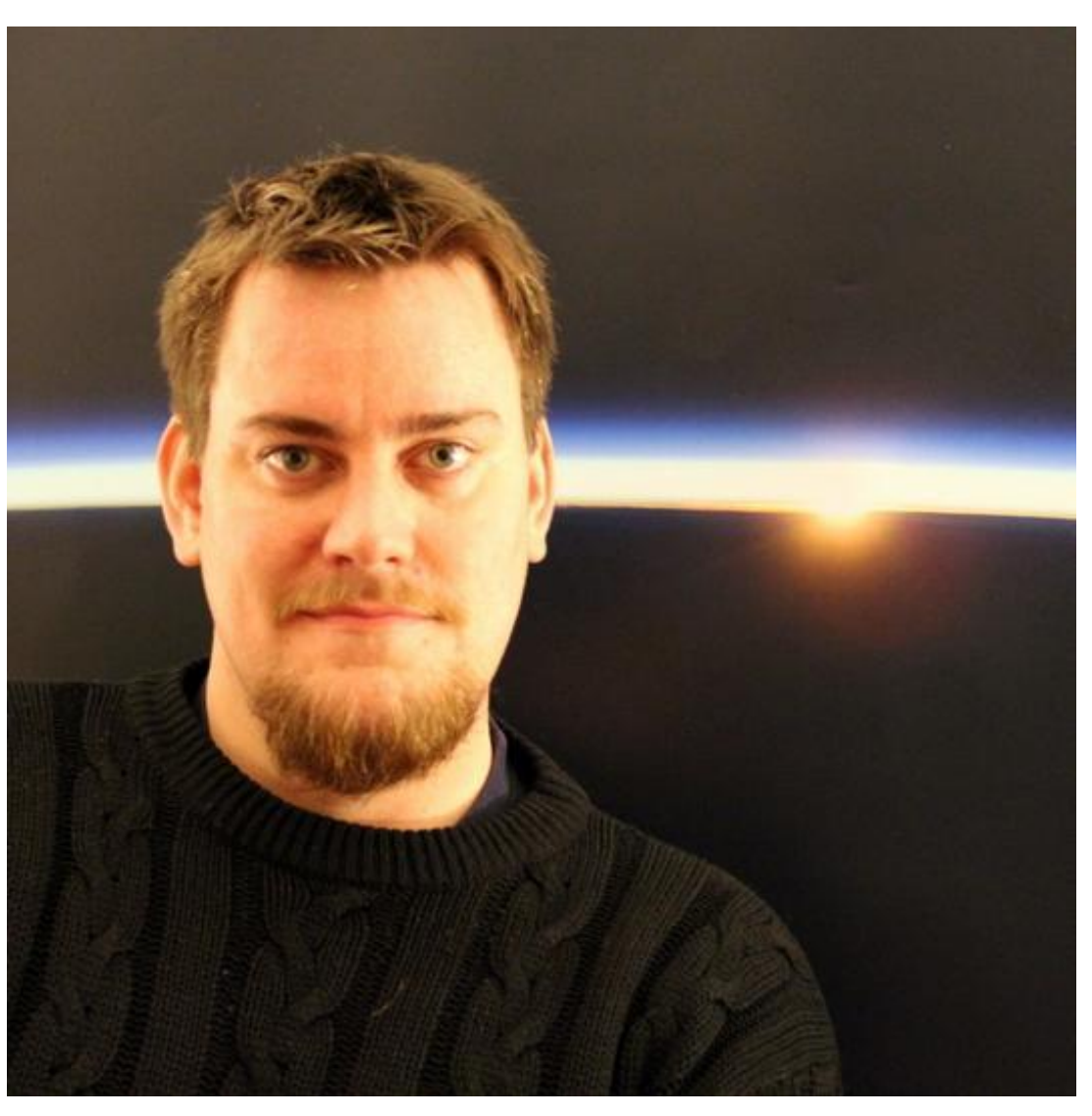

Rasmus Handberg, creator and long-time maintainer of the KASOC and TASOC platforms.

Manager at the Stellar Astrophysics Centre (SAC) in Aarhus. In addition to continuing the development of TASOC (with one of the editors, MNL, having had the privilege of being both his office mate and resident guinea pig for new features), Rasmus also co-chaired the TASC Coordinated Activity on 'TASC Data Products' and played an important role in supporting the target selection work for both *Kepler* and TESS.

KASOC provided the community with a comprehensive set of tools spanning data access, internal communication, and collaborative infrastructure. On the data side, mission data could be downloaded with extensive search options, the input catalogues (KIC/EPIC/TIC) could be queried with direct links to available data products, published papers, and SIMBAD entries for any given star, and working groups could host and share their own processed data products. For communication and community building, members could access conference material from KASC/TASC workshops, share publications for internal review, and find contact information and mailing lists for all consortium members. With TASOC, these foundations were extended by dedicated wiki pages for each working group—improving coordination of projects and target selection—and, during the Covid-19 pandemic, a system for advertising online talks and events.

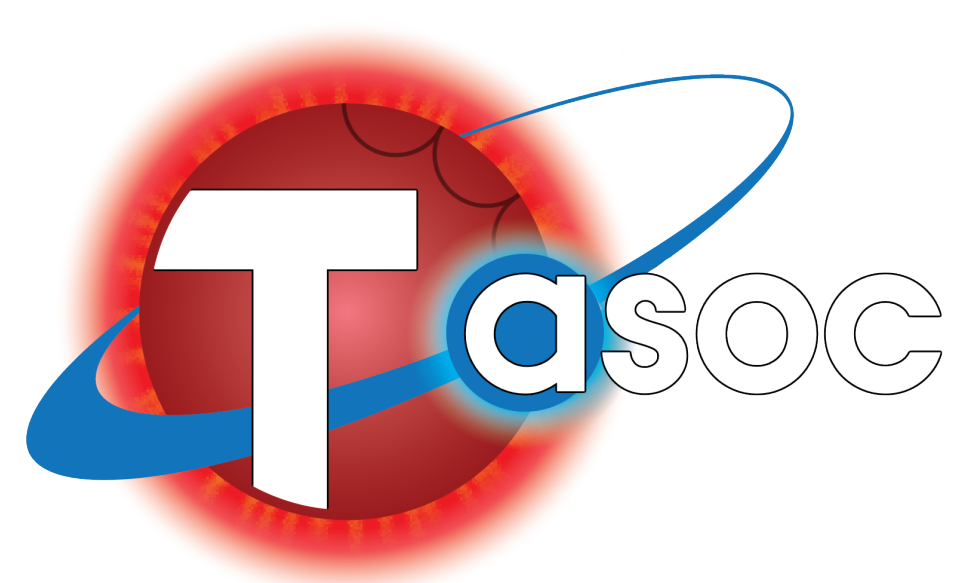

Logo of the TASOC database, inspired by the official logo for the TESS mission.

In short, while the formation of KASC signified the start of a new collaborative era in asteroseismology on paper, it was the platforms of KASOC and TASOC that truly provided the glue binding the community together.

Sadly, during the summer of 2023, Rasmus fell ill and could no longer continue his work and support of the databases. At the time of writing, the databases are therefore not actively supported, and many features no longer work as intended. However, the importance of the platforms and the functions they provide—still allowing advertising of new papers to the community, sharing information on upcoming conferences and job openings via the hosted mailing lists, and providing an entry point for postdocs and PhD students to take part in collaborations—is well acknowledged by both the Aarhus team running the sites and the TASC Steering Committee, and plans are in motion for a revitalisation of the platforms.

# III
# Conferences

## The KASC/TASC Workshop Series

The annual KASC/TASC Workshop Series—so named in its early years, when the meetings were smaller in scale—has served as the main scientific forum for the international asteroseismic community. Over time, these meetings have grown into full-scale conferences. They provide a structured yet dynamic framework for presenting new results, consolidating methodological advances, and defining strategic scientific directions. Throughout the years, the workshops have played a decisive role in consolidating asteroseismology as a cornerstone discipline within stellar astrophysics, while simultaneously strengthening its connections to exoplanetary science, Galactic archaeology, and stellar population studies.

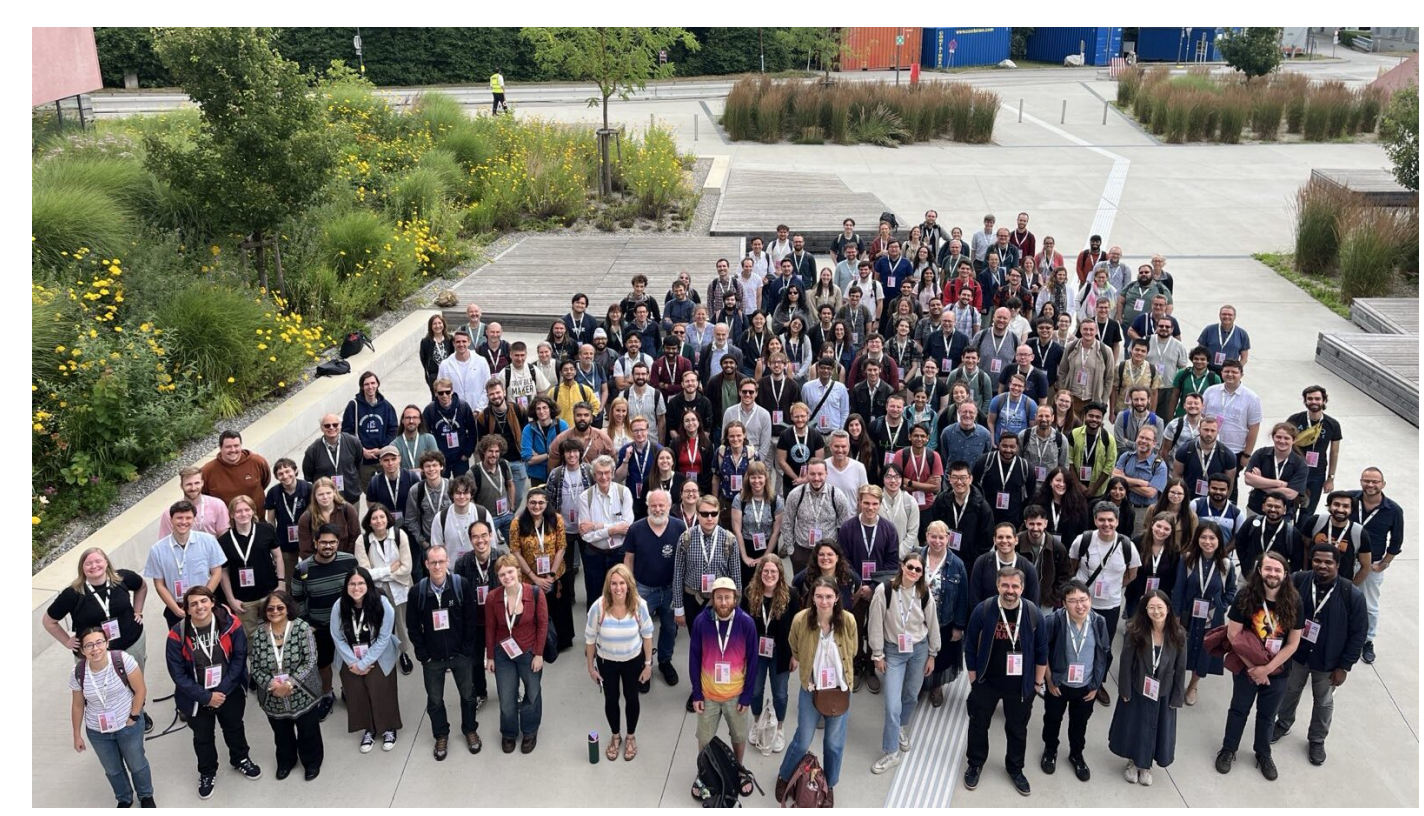

Group photo taken during the TASC9/KASC16 Workshop (Vienna Woods, Austria, 2025).

Since 2007, the KASC/TASC workshops have been organised annually at major international research centres. A particularly important milestone in the evolution of the series was the 2014 joint CoRoT Symposium 3 – KASC7 Workshop, which marked a convergence between the CoRoT and *Kepler* asteroseismic communities at a time when space-based photometry was reshaping stellar astrophysics. This meeting also recognised the earlier community-building efforts of the CoRoT consortium, whose biannual "CoRoT Weeks" had brought together researchers working on different classes of pulsating stars since the early 2000s. This was followed, in 2015, by the first joint KASC/TASC workshop (TASC1/KASC8), reflecting the natural extension of the established *Kepler* asteroseismic consortium framework to encompass the forthcoming TESS mission within a unified conference structure. Together, these developments reinforced the evolution from mission-specific activities toward a broader, mission-agnostic collaboration. The group photo shown alongside this text, taken during the TASC9/KASC16 Workshop (Vienna Woods, Austria, 2025), illustrates the scale and continuity of this international community. Below is a chronological overview[1]:

- **2007** — First KASC Workshop, Paris, France.
  `http://www.ias.fr/kepler`
- **2008** — Second KASC Workshop, Aarhus, Denmark.
- **2010** — Third KASC Workshop, Aarhus, Denmark, "*Kepler* Asteroseismology in Action".
- **2011** — Fourth KASC Workshop, Boulder, USA, "From Unprecedented Data to Revolutionary Science".
- **2012** — Fifth KASC Workshop, Balatonalmádi, Hungary, "Extending the *Kepler* Mission: New Horizons in Asteroseismology".
- **2013** — Sixth KASC Workshop, Sydney, Australia, "A New Era of Stellar Astrophysics with *Kepler*".
- **2014** — CoRoT Symposium 3 – KASC7 Workshop, Toulouse, France, "The Space Photometry Revolution".
  `https://corot3-kasc7.sciencesconf.org/`
- **2015** — TASC1/KASC8 Workshop, Aarhus, Denmark, "Space Asteroseismology: The Next Generation".
  `https://phys.au.dk/sac/past-events/the-kasc8tasc1-workshop-2015/`
- **2016** — TASC2/KASC9 Workshop – SPACEINN & HELAS8 Conference, Angra do Heroísmo, Terceira (Azores), Portugal, "Using Today's Successes to Prepare the Future".
  `https://www.iastro.pt/research/conferences/spacetk16/`
- **2017** — TASC3/KASC10 Workshop, Birmingham, UK, "TESSting Stellar Astrophysics".

[1] Web links are provided where active conference pages or institutional archive pages are currently available.

- **2018** — TASC4/KASC11 Workshop, Aarhus, Denmark, "First Light in a New Era of Astrophysics".
  `https://conferences.au.dk/tasc4`
- **2019** — TASC5/KASC12 Workshop, Boston, USA.
  `https://web.mit.edu/tasc5`
- **2022** — TASC6/KASC13 Workshop, Leuven, Belgium, "Asteroseismology in the Era of Surveys from Space and the Ground: Stars, Planets, and the Milky Way".
  `https://fys.kuleuven.be/ster/events/conferences/2020/tasc6`
- **2023** — TASC7/KASC14 Workshop, Honolulu, Hawaii, USA.
  `https://tasc.ifa.hawaii.edu/`
- **2024** — TASC8/KASC15 Workshop, Porto, Portugal.
  `https://www.iastro.pt/research/conferences/tasc8-kasc15/`
- **2025** — TASC9/KASC16 Workshop, Vienna Woods, Austria.
  `https://tasc9-kasc16.ista.ac.at/`
- **2026** — TASC10/KASC17 Workshop, Aarhus, Denmark, "A Nod to the Past with a View to the Future".
  `https://conferences.au.dk/tasc10-kasc17`

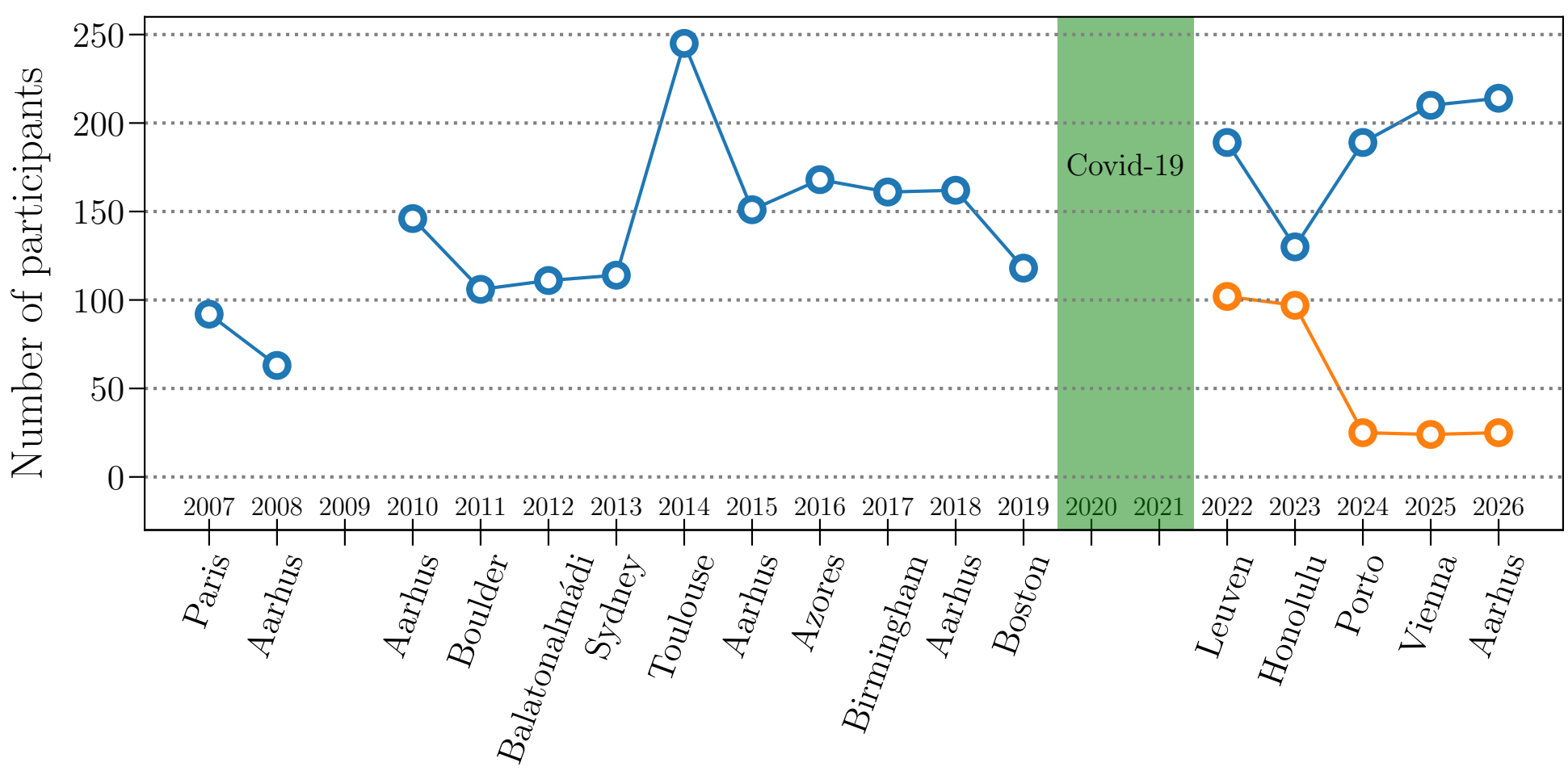

**Figure 2**: Evolution in the number of participants attending the KASC/TASC workshops over time. Blue markers indicate in-person participation, while orange indicates registered online participation. For online participation, we note that many of the workshops before 2019 also allowed online attendance, but registration records from these have not been available.

Over nearly two decades, the KASC/TASC Workshop Series has provided a stable yet evolving framework for coordinating international efforts in space-based asteroseismology.

Beyond enabling rapid dissemination of results and fostering cross-institutional collaborations, recent editions have further strengthened their community-building mission through dedicated mentor sessions to support early-career researchers, as well as discussion sessions addressing equity, diversity, and inclusion within the community, including LGBTQIA+ perspectives in astronomy. These efforts have been complemented by support from the White Dwarf Research Corporation (`https://wdrc.org/`), founded and led by Travis Metcalfe, which has facilitated the participation of early-career researchers, including by sponsoring registration fees.

As the field transitions toward exploiting long-time-baseline datasets and prepares for next-generation facilities, the workshop series continues to serve as a strategic platform for defining scientific priorities. Its sustained success reflects both the scientific vitality of asteroseismology and the collaborative structure established by the KASC and TASC communities.

Figure 2 shows the evolution in the number of participants attending the KASC/TASC workshops. Only in 2009 (the year of *Kepler*'s launch) and in 2020–2021 during the Covid-19 pandemic were there no workshops. Since the early workshops focused on preparing for *Kepler* data, attendance has steadily doubled to $\sim$200 participants.

## Summer schools

Two editions of the KASC/TASC Workshop Series were followed by advanced international summer schools, extending the conferences' community-building mission into structured training initiatives. These schools were designed to equip MSc and PhD students, as well as early-career researchers, with theoretical and practical expertise in asteroseismology and related fields, including exoplanetary science and stellar population studies. Through lectures and hands-on tutorials, participants were trained to extract and interpret stellar properties from space-based photometry and complementary ground-based spectroscopy, preparing them to lead the scientific exploitation of current and forthcoming facilities.

Both schools emphasized the methodological and scientific advances in space-based asteroseismology achieved during missions such as CoRoT, *Kepler*, K2, and TESS, and looked ahead to the scientific landscape shaped by PLATO, *Roman*, Earth 2.0, and HAYDN while highlighting the role of complementary spectroscopic networks and surveys.

- **2016** — *Asteroseismology and Exoplanets: Listening to the Stars and Searching for New Worlds*
  IVth Azores International Advanced School in Space Sciences
  Horta, Faial (Azores), Portugal
  `https://www.iastro.pt/research/conferences/faial2016/`

➢ **2024** — *Porto Summer School on Asteroseismology: From Pixels to Stellar Ages*
Azurara, Vila do Conde, Portugal
`https://www.iastro.pt/research/conferences/pssa2024/`

# IV
# Personal Testimonials

In what follows, we present personal reflections on the early days of KASC and TASC from several individuals who played central roles in the formation of the consortia.

## Jørgen Christensen-Dalsgaard

*Professor Emeritus, Aarhus University, Denmark*

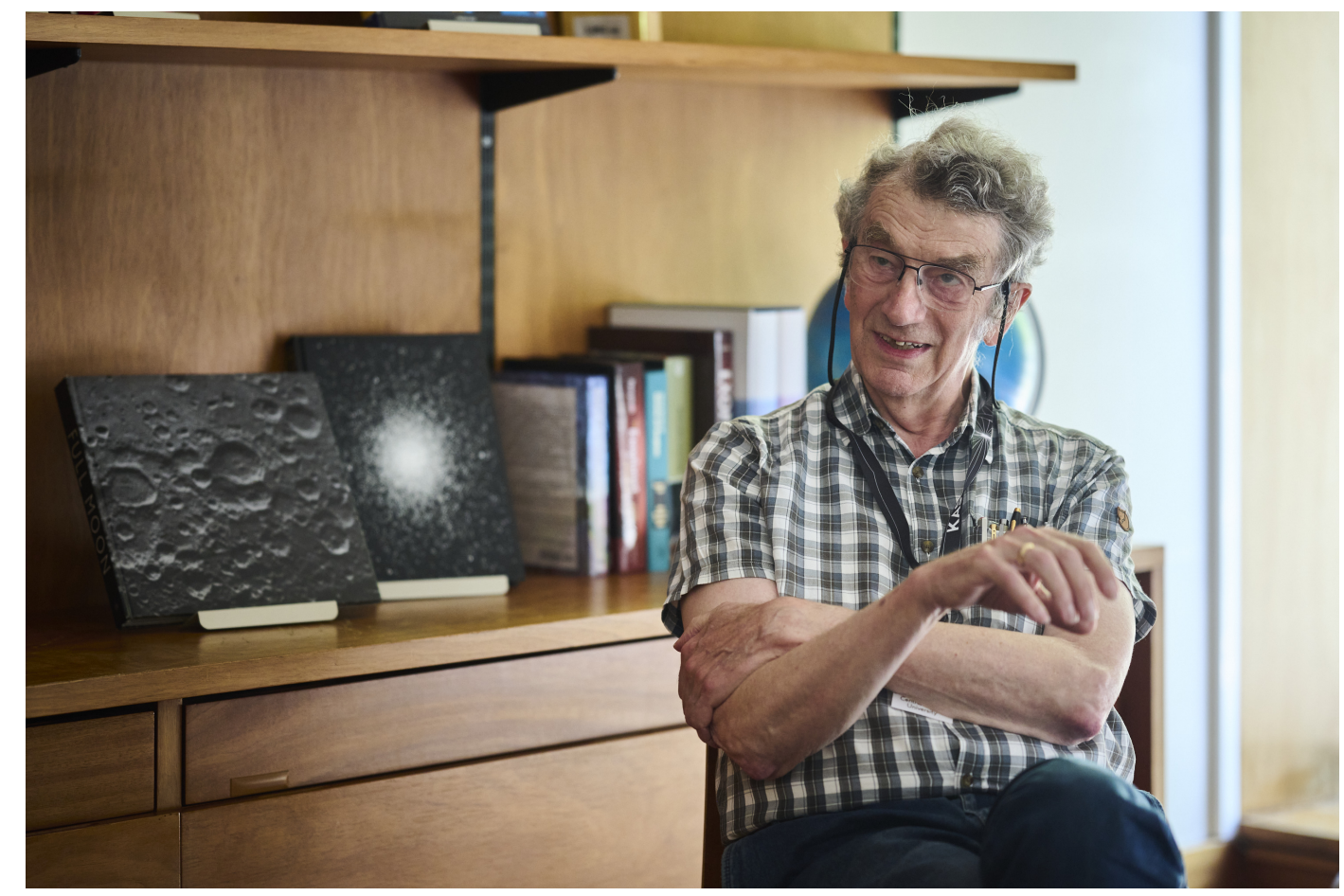

Here, I present a personal history of the way the astronomy group at Aarhus University got involved in the *Kepler* project, leading to the establishment of KASC. The group was heavily involved in potential space missions for asteroseismology, including the ESA *Eddington* proposal and the Danish-led *Rømer* satellite project with the MONS investigation dedicated to asteroseismology. *Rømer* reached a high level of design and preparations for science, but failed to obtain funding in the 2003 Danish national budget. Also, *Eddington*, for a while a "reserve mission" in ESA's programme, was finally cancelled in 2003. Even so, the group had substantial expertise in the planning for space asteroseismology.

This provided the basis for the involvement in the *Kepler* mission. There was already collaboration with Tim Brown and Ron Gilliland, who were closely involved in *Kepler*, but the real kick-off came with a visit to Aarhus by the *Kepler* Deputy PI, Dave Koch, in September 2003. Dave had seen the launch of *Rømer* announced on the (woefully out of date) *Rømer* web pages and was interested in our experience with the mission. He was already in Europe and, at short notice, arranged a visit to Aarhus to give a seminar on *Kepler*. The result of discussions with Hans Kjeldsen and me was an invitation to participate, in some manner, in the mission. The precise manner of participation was initially unclear. I had a meeting with the *Kepler* PI, Bill Borucki, in Boulder in August 2004, together with Tim Brown. In a mail following this meeting Borucki initially doubted whether I could be included as CoI so relatively late in the project. At least this would require approval from NASA HQ. Alternatives were involvement through the Participating Scientist Program (PSP), or as Guest Observer (GO). However, Ron Gilliland at the time was arguing strongly for the CoI option.

Following these initial discussions, I attended the *Kepler* Science Team meeting in Boulder in October 2004 and made a presentation on asteroseismology. Following this, the required activities were set in motion for me to obtain CoI status, involving Carl Pilcher at NASA HQ and also supported by the grant for the Danish AsteroSeismic Centre (DASC) that we received from the Danish Independent Research Fund. At least preliminary CoI status was in place for the May 2005 Science Team Meeting at the STScI. Hans Kjeldsen also attended the November Science Team Meeting of that year, presenting two talks, the latter describing specifically the planned activities in Aarhus. Here he also raised the question: 'We may want to expand the group that are involved in preparing the p-mode seismology. How should we handle this?'

An important issue was how access to the asteroseismic data would be arranged. The mission operated with a limited number, up to 500, of targets observed at one-minute cadence, while the bulk of the targets were to be observed at 30-minute cadence. The asteroseismic investigation was seen as being mainly concerned with the high-cadence data, which would allow seismology of main-sequence stars. The original proposal from Borucki was that this could be done through the PSP or GO programs. This involved discussions with Gilliland, who already had assured direct access to the data. In a letter to Borucki in March 2006, Gilliland argued strongly that the Aarhus group should have direct access to the short-cadence data. It was also becoming increasingly obvious that the amount and quality of the expected data were far broader than could reasonably be handled by this small group. Out of this grew the idea of the *Kepler* Asteroseismic Science Consortium (KASC) and the *Kepler* Asteroseismic Science Operations Centre (KASOC) based in Aarhus, to ensure the participation of a broad scientific community in the analysis of the data. This was spelled out in a document, led by Hans Kjeldsen, defining the *Kepler* asteroseismic Key Programme. There were issues with the organization of control of the data access and accidental inclusion of data that could lead to uncontrolled early discoveries of exoplanets. The result of these discussions was a draft agreement in December 2006, intended also to include a NASA signature, laying out the details of the organization, including KASC and KASOC. However, it was felt that an international agreement at NASA level would be too complicated, potentially involving also the US State Department. As a result, the agreement draft was transformed into a Letter of Direction from Borucki as *Kepler* PI, to the *Kepler* Asteroseismic Investigation represented by Gilliland (see § A), and laying out the conditions for the asteroseismic activities, including the formation and operations of KASC.

On this basis, a preliminary list of potential members of KASC was set up and an invitation letter sent out by the KASOC secretary Danijela Jelicic at the end of May 2007 (see § B). Also, it was arranged to organize the first KASC Workshop at IAS, Paris, in October 2007. There was immediate interest in participating in the project in this manner: at the end of November 2007 Danijela noted in a mail to Pamela Marcum, NASA, that KASC had around 260 registered members, and the number later grew to around 500. In connection with the workshop at IAS, a meeting of the KASC Steering Committee was held, with detailed discussions of various aspects of the project, including data-analysis pipelines and procedures for target selection, requiring scientific justification for each target. This also included a first discussion of setting up working groups for the different target types with a view towards target selection and collaborative analysis of the data; these were later formalized and became central to the asteroseismic work on *Kepler* data.

Following this first KASC meeting, there was an ongoing analysis and discussion of the procedures for selecting targets for the *Kepler* short-cadence slots, with a view towards asteroseismology and including the prospects for characterizing planet-hosting stars. This included the identification of an initial survey period covering a range of different relevant targets. On this basis, specific targets would be selected for longer observing sequences. These discussions continued in the second KASC Workshop in Aarhus in June 2008. Here, there were also presentations of the status of the project, including the *Kepler* Input Catalogue, which was an important tool for target selection. Furthermore, there were presentations of the potential of *Kepler* data for a variety of different stellar types.

The first call for *Kepler* asteroseismic targets was issued with a deadline of September 2008. For the short-cadence observations two types of targets were considered: survey targets to be observed for relatively short periods to determine the possibilities of *Kepler* for a broad range of stellar types; and dedicated targets to be observed typically for the rest of the duration of the nominal mission. To increase the coverage of the survey, it was decided that the survey targets would be observed for 30 days each, which was expected to be sufficient to give a realistic impression of the quality of the observations. Dedicated targets would in general be selected from stars that had been surveyed on the basis of the results of the survey. On the basis of these proposals, the KASC Steering Committee prepared the first list of targets to be transmitted to the *Kepler* Science Office, covering the first year of observations. Additional samples of stars observed at long cadence, both comprising red giants, turned out to be of the greatest importance to asteroseismology. The first was a sample selected as astrometric reference and a second, fortuitous, sample consisted of red giants misclassified as dwarfs and hence included in the exoplanet programme.

In preparation for the analysis of the first *Kepler* data, KASC was organized into working groups with responsibility for separate types of targets. These were also charged with the preparation of the first key papers based on the early data. The basis for the formation of the working groups was letters of intent provided with an initial deadline of the end of 2008. The submission of a letter of intent (§ D) and signing of a non-disclosure agreement (see § C) to avoid unwarranted release of possible exoplanet detections were requirements for a KASC member to be associated with a working group and to be provided with a password that gave access to the data on KASOC. The result of this exercise was the formation of the following working groups:

1. Solar-like p-mode Oscillations
2. Oscillations in Clusters
3. Beta Cephei Stars
4. Delta Scuti Stars
5. roAp Stars
6. Slowly Pulsating B-stars
7. Cepheids
8. Red Giants
9. Pulsations in Eclipsing Binaries
10. Gamma Doradus Stars
11. Compact Pulsators
12. Mira Stars
13. RR Lyrae Stars
14. RV Taurus Variables

Most working groups were divided into subgroups with specific responsibilities (time-series analysis, modelling, ground-based follow-up, etc.), and with chairs of the working group and the subgroups. This organization turned out to work remarkably well to ensure a very collaborative and productive analysis of the asteroseismic data. A crucial requirement for the work of the KASC working groups was the establishment of KASOC to make the data available and to facilitate the collaboration of the working-group members and the publication of the results. This clearly had to be in place before the first release of *Kepler* data. Admittedly, the work on setting up KASOC got off to a rather late start, but fortunately Rasmus Handberg, then a PhD student supervised by Hans, accepted to lead the development and brilliantly succeeded in having a first version ready in time, with support from Søren Frandsen. KASOC, further developed in the following years by Rasmus and others, has been invaluable for the very successful work of KASC.

Together with many members of the *Kepler* community, I was present at Cape Canaveral for the launch on 6 March 2009, my first experience of a rocket launch after having followed the launches (and failures) from Cape Canaveral since I was a young boy. An unforgettable experience with the night-time launch and the confirmation that the spacecraft was alive and had entered the planned orbit. After a 10-day commissioning period, *Kepler*'s observations were organized in quarters, separated by a spacecraft roll to keep the solar panels properly pointed towards the Sun, and with a one-month first quarter (Q1). Long-cadence commissioning data were released on 26 August, announced in the first news item on KASOC.

The Q1 long-cadence data were released on 14 September and the Q1 short-cadence data 23 November. These early data fully demonstrated the exquisite quality of the *Kepler* data and led to immediate work on data analysis. I led an analysis of the known, from ground-based transit observations, exoplanet-host star HAT-P-7, together with Hans, Ron Gilliland, Tim Brown and others. Here, a spectrum of individual frequencies was obtained, allowing a detailed asteroseismic analysis with techniques later used in extensive work on exoplanet hosts and the so-called Legacy Sample of stars. This was published in March 2010, in a special issue of Astrophysical Journal Letters dedicated to *Kepler* performance and early results. Out of 24 papers in that issue, 10 were related to KASC activities, covering several different types of variable stars.

A landmark event for KASC was the third KASC Workshop, which took place in June 2010 in Aarhus. This was probably the first large international meeting dealing with *Kepler* asteroseismic data and demonstrated the incredible range and quality of the data. The programme covered essentially all the relevant stellar types, with presentations of light curves and oscillation spectra of a nature never seen before, the physical origin of which was still a puzzle and extensively discussed. The workshop also provided an opportunity for meetings of the individual working groups, in the light of the first data, setting the pattern for the following yearly KASC workshops. In connection with the workshop, a meeting of the *Kepler* Science Team was also held in Aarhus.

I shall stop my personal description of our involvement in the *Kepler* project and KASC at this key point. My continued work related to KASC has included further model fitting to *Kepler* observations. Also, I have increasingly worked on the properties of red-giant structure and oscillations, an area where the *Kepler* data for me have perhaps delivered the most striking results.

## William (Bill) Borucki

*Kepler PI, NASA Ames Associate (retired), USA*

### Origins of the *Kepler*– Asteroseismology collaboration

After the *Kepler* Mission was approved for development, two members of the *Kepler* Team, Ron Gilliland and Timothy Brown, who were active in asteroseismic research, pointed out that the very high precision photometric observations by the *Kepler* photometer could make valuable contributions to astrophysics research. Because the Mission would make nearly continuous photometric measurements of thousands of stars for periods of months to years, it would be possible to determine the characteristics of these stars and their evolutionary state. Thus, cooperation between the *Kepler* Science Team and the international community of asteroseismologists would make possible breakthroughs in astrophysical and exoplanet science.

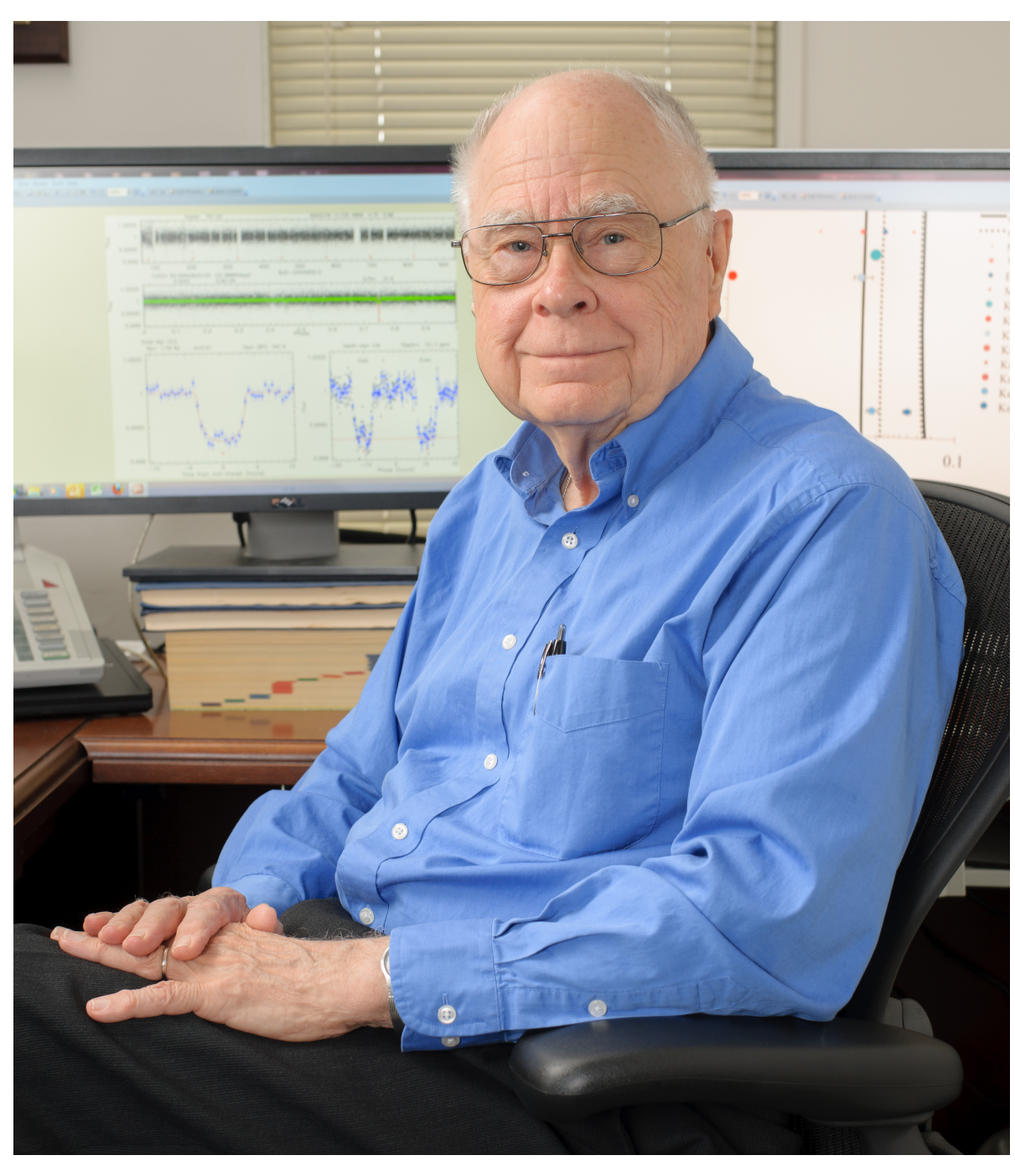

Sharing the *Kepler* photometric data would address the needs of both stellar astrophysicists and exoplanet scientists. The first group was interested in the intrinsic stellar properties, processes, and evolution, while the second group needed the determination of stellar characteristics, such as the size of the star, to determine the size of the planets. For the cooperation to be fruitful, careful consideration of the choice of targets, the measurement cadence, and the duration of the observations was necessary.

### Formation of KASC and the Letter of Direction

The KASC consortium was shaped by a "Letter of Direction" between the *Kepler* PI William Borucki and Jørgen Christensen-Dalsgaard at Aarhus University. Usually, NASA HQ would arrange such an agreement through the NASA International Program Office, but they recognized that foreign scientists would not need to be involved in the technical design of the Mission, so a simple agreement with the PI and a scientist representing the consortium was sufficient.

Ron Gilliland arranged a meeting in Europe where I could meet with members of the European asteroseismology community. We discussed the types of measurements that were of most interest to the astrophysics community and the types and brightness of main-sequence stars that were most useful to exoplanet characterization. We found that collaboration would provide substantial benefit to both groups. The *Kepler* project would observe a limited number of giant stars of special interest to asteroseismologists, and in turn the asteroseismic community would characterize the much smaller stars that were of special interest to the exoplanet community. To accommodate the asteroseismology measurements, it was necessary to increase the measurement cadence beyond that needed to detect planetary transits. Therefore, a compromise needed to be found because the higher cadence rate for the asteroseismic targets adversely affected the total number of stars that could be observed due to the limitation for onboard data storage and for communication bandwidth.

I was the *Kepler* PI and it was my responsibility to guide the science in accomplishing the major goals of the Mission; in particular, to determine the occurrence frequency of exoplanets and especially of near-Earth-size planets in and near the habitable zone of solar-like stars in our galaxy. To accomplish those goals, it was important to obtain data that allowed accurate estimates of the size, mass, age, and temperature of the stars. Such data both characterize the host star and allow the deduction of planet size and whether a planet is in the habitable zone of its star. I recognized that the development of KASC would make an important contribution to the Mission goals.

A major challenge was the cost of the increased complexity of the data acquisition as well as the cost of supporting scientists who would be doing asteroseismic research rather than exoplanet research. The problem was solved by a compromise with NASA HQ that allowed the Mission to acquire the data needed by the asteroseismologists while avoiding payment of their salaries and research expenses. Thus the Mission provided the necessary data while our colleagues conducted their research with funding from their sponsoring agencies.

## Scientific impact and major achievements

There are two science fields to be considered; exoplanet science and stellar astrophysics. In the field of exoplanet science, major contributions by the asteroseismologists were the accurate stellar sizes, temperatures, and ages for a portion of stars with planets. With respect to stellar astrophysics, long-term p- and g-mode observations allowed new information on the structure and characteristics of stars, especially oscillating stars and red giants. Internal differential rotation for red giants indicated the need for further development of theories of the evolution of giant stars. Another result was the observation of binary stars on highly-eccentric orbits that showed the onset of high-amplitude waves excited at periapsis and their subsequent damping as the stars separated. Special highlights were the determination of the size of some planets to an accuracy of a few percent and the capability to distinguish hydrogen-shell burning stars from those burning helium in their cores in stars that appeared identical based on exterior observations.

On a personal level, my most memorable experiences occurred when I attended presentations of the determination of the characteristics of stars deduced from *Kepler* asteroseismology measurements. The actual measurements of stellar properties, rather than crude estimates based on spectral type, greatly increased the accuracy of planet size and the incident radiant flux needed to determine if the planet was in the habitable zone.

### Legacy and future perspectives

The most important legacy of KASC was the plethora of new knowledge about both stellar structure and exoplanet properties from the productive collaboration of international teams in two different science disciplines. Based on the data from the *Kepler* and TESS missions, and on future missions with advanced capabilities, I expect that more complete and accurate data covering an even greater variety of stars will lead to new discoveries. These data will continue to promote the very vigorous and dynamic field of stellar astrophysics and will bring many young investigators into the field.

## Hans Kjeldsen

*Professor, Aarhus University, Denmark*

One of the things I remember best from the early *Kepler* days is the first time we saw the real data. For years, we had talked about what *Kepler* might do for asteroseismology. We had made predictions, discussed observing plans, argued about targets, and prepared ourselves for photometry of a quality we could never get from the ground. But it is one thing to say that something should work. It is quite another to sit there and see it work.

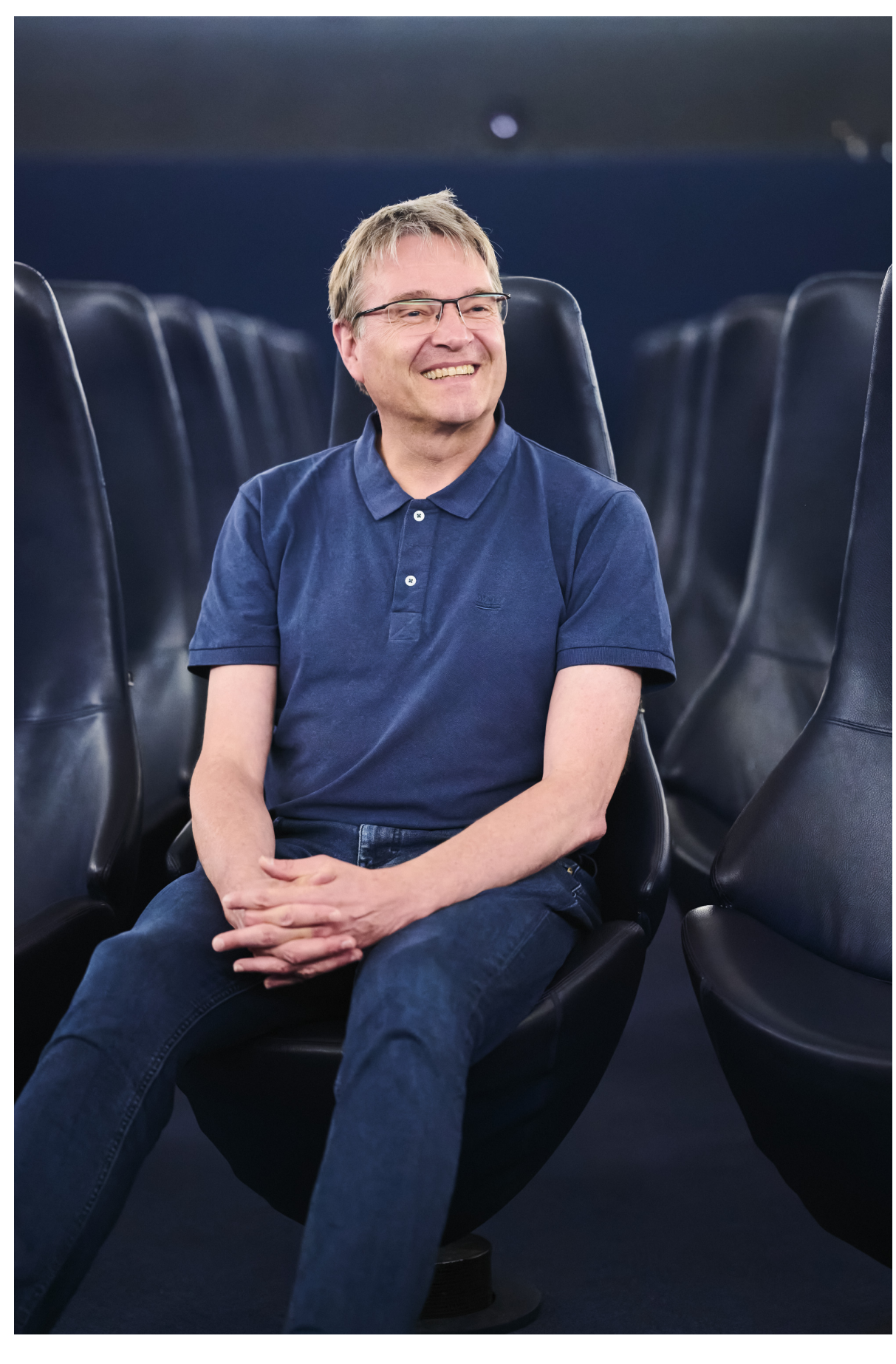

The first light curves and power spectra were a very special moment. I think many of us had expected the data to be good, but still, the reality was almost shocking. The oscillation signals were not just barely visible after difficult analysis. They were very clear. In some stars, the power spectra almost looked like textbook examples. Features that had been at the edge of what we could do from the ground were suddenly there with a precision and continuity we had only dreamed about. For me, this was the moment when it became obvious that *Kepler* would not just give us better data. It would change the whole field. It was not only that we could study individual stars better. It was then that we could begin to think about asteroseismology as a large-scale tool.

KASC came out of that expectation, but also out of many years of earlier work. The community was already connected through space missions, mission proposals, ground-based campaigns, and smaller collaborations. The Canadian MOST mission and the French-led CoRoT mission had shown what space photometry could do. CoRoT was especially important, because it showed that one must be ready before the data arrive. Data do not become science by themselves. The people, the tools, the target lists, and the collaborations must be in place. The Danish *Rømer* satellite project, with MONS as its asteroseismic payload, had also made us in Aarhus think very practically about targets, observing modes, data analysis, and how one might organise a community around a space mission. ESA's *Eddington* mission, SONG, and several ground-based campaigns were also part of this background.

For the Aarhus group, the road to *Kepler* was a mixture of disappointment and good fortune. Under the leadership of Jørgen Christensen-Dalsgaard, we had spent many years working towards space asteroseismology. The cancellation of *Rømer* was a major disappointment, and the cancellation of *Eddington* was another. Both missions would have produced wonderful data. But, looking back, they also left us with ideas, experience, software concepts, target-selection plans, and a network of collaborators. So, when *Kepler* became a real possibility for asteroseismology, we were not starting from nothing.

The opportunity became concrete when Jørgen was invited to join *Kepler* as a Co-Investigator, with responsibility for helping to make scientific use of the short-cadence observations. At that time, the short-cadence capacity was not the main data product of the exoplanet programme, but for asteroseismology, it was essential. Jørgen asked me to take on much of the practical work of developing the asteroseismic programme. From 2005, I attended the *Kepler* Science Team meetings and travelled frequently to the USA, where I got to know Bill Borucki, Dave Koch, Natalie Batalha, David Latham, Jack Lissauer, Jason Rowe, Steve Bryson, Jon Jenkins, Steve Howell, Sara Seager, and many others in the *Kepler* project. Together with Ron Gilliland and Tim Brown, we began to work out what could be done.

Some things became clear very quickly. The precision would be good enough to detect solar-like oscillations in many stars. The number of possible targets would be far too large for a small group to handle. And the science would not be limited to one type of star. If *Kepler* were to be used properly for asteroseismology, we needed expertise from across stellar astrophysics. It was not just a question of which stars we wanted in Aarhus. The question was how to involve the broader community in choosing and using the targets in a sensible and fair way.

That was the practical origin of KASC. We needed an international structure that could do two things at the same time. It had to give a broad scientific community access to *Kepler* asteroseismology, and it had to respect the rules and sensitivities of the *Kepler* exoplanet mission. Two issues were especially important. First, access to the data had to be organised so that KASC would not be seen as a back door into early exoplanet discoveries. Second, the data, target lists, working groups, and publications had to be coordinated well enough that a large community could work together.

At the same time, it became clear in the *Kepler* Science Team discussions that asteroseismology could help the main exoplanet mission. Planetary radii depend directly on stellar radii, and asteroseismology offered one of the best ways to determine stellar radii accurately for suitable stars. So, the asteroseismic programme was not just an extra science case attached to *Kepler*. It could also improve the interpretation of the planets *Kepler* would find. This link between stellar physics and exoplanet science became an important argument in the discussions.

The formal arrangement was eventually written into the Letter of Direction. After many discussions, the best solution was not a small closed asteroseismic team, but an open international consortium with clear rules. This defined two urgent tasks: opening membership of KASC and forming KASOC at Aarhus University. I was asked to lead much of the organisational work as project scientist for KASC, while the international steering structure took shape around the science priorities and policies.

KASOC was essential. Without a working data and coordination centre, KASC would have remained a good idea, but not a functioning consortium. Rasmus Handberg played a central role in making this possible. Through the KASOC website and the systems behind it, the consortium could handle data access, target information, working-group organisation, paper coordination, and all the daily details that large collaborations depend on. These things are not always visible in the final scientific papers, but without them, the papers would not have happened.

The target selection was organised through working groups covering different classes of stars. This was partly inspired by earlier experience from MONS, CoRoT, and other projects, but it also followed naturally from the scale of *Kepler*. No single group had the expertise to judge the best targets across the HR diagram. We needed people who knew solar-like oscillations, red giants, compact pulsators, delta Scuti stars, gamma Doradus stars, massive stars, eclipsing binaries, clusters, and many other kinds of variable stars. The working groups were the practical way to bring all that expertise into the target selection and later into the analysis.

The early phase also had its difficult moments. The most sensitive issue was the possibility that asteroseismic data might contain exoplanet transits. The *Kepler* mission had to protect its planet discoveries, and KASC had to accept special rules for data access. This led to non-disclosure agreements and, for a while, to a transit-removal filter. The filter was meant to solve a real problem, but it also created problems for the asteroseismic analysis. It was one of the more frustrating parts of the early collaboration. Still, I think it should be remembered as part of a negotiation between two scientific communities that both had good reasons for their concerns. In the end, we found a way through it.

One of the most surprising *Kepler* legacies came from the red giants. Before the *Gaia* space mission, we did not have accurate parallaxes and, therefore, distances to stars in the *Kepler* field. With only apparent brightness and colour, it was not always possible to separate main-sequence stars from evolved red giants. As a result, many red giants entered the *Kepler* target lists. At the time, some of them may not have looked like the most obvious targets. Had we already had *Gaia*-quality parallaxes, I am sure we would have removed many of them, especially since the aim of the *Kepler* target selection was to focus mainly on solar-like dwarfs and the best exoplanet targets.

Fortunately, we did not. The red giants became one of *Kepler*'s greatest gifts to astrophysics. In the beginning, one could almost say that they polluted the samples, at least from the point of view of selecting clean samples of main-sequence stars. But that "pollution" turned out to be extremely valuable. The red giants were bright enough, numerous enough, and oscillating beautifully, so they became a new way into stellar astrophysics. Their oscillations were detected in large numbers, and it quickly became clear that these stars were a treasure in the data.

This changed what we could do with asteroseismology. With red giants, we could move from studying individual stars to studying stellar populations. From their oscillations, we could determine radii, masses, evolutionary states, and eventually ages for very large samples. This made it possible to connect stellar interiors with the history of the Milky Way. The red giants became clocks and tracers. They allowed us to ask where different stellar populations formed, how they migrated, and how the Galaxy built up over time. What could have looked like a population of less useful targets became one of the foundations for Galactic archaeology with asteroseismology.

For me, this is one of the strongest lessons from *Kepler*. When a mission opens a new observational window, one should be careful not to select only the targets that fit what one already expects to be important. Sometimes, the apparently secondary objects contain the real discovery space. In the case of *Kepler*, the lack of precise distances before *Gaia* helped preserve a red-giant sample that later became fundamental for using asteroseismology as a tool for broader astrophysics. This has consequences not only for stellar evolution, but also for our understanding of Galactic structure, stellar populations, and the age scale used in many areas of astrophysics. In this way, *Kepler* asteroseismology became relevant far beyond the original stellar-physics questions, reaching into Galactic evolution and even into the broader framework in which stellar ages are used in cosmology.

KASC was built on the idea that the data should be used broadly and collaboratively. That sounds simple, but in practice it requires patience, rules, trust, and a willingness to share credit. It also requires care for young researchers, whose projects can easily disappear when a flood of new data arrives, and everyone is excited. I think KASC managed this balance quite well. The working groups gave structure, but the scientific energy came from the members. The result was a consortium that was sufficiently organised to function, but at the same time open enough that new ideas could grow.

Another very important part of KASC, and later TASC, was the yearly workshops. These meetings were much more than places where we presented results. From the beginning, the KASC and TASC workshops were where the community met, planned, argued, learned, and developed the programme together. We met in different places around the world, and these meetings became part of the culture of the field. They were important for the steering committees and working groups, but they were just as important for students and young researchers. Many people were introduced to asteroseismology and international collaboration through these workshops. I think that has been one of the most valuable parts of the consortia.

Through the steering committees, working groups, and workshops, I also had the pleasure of strengthening many collaborations that were already developing, and starting new ones. The connections with the groups in Sydney, Birmingham, Porto, MIT, and many other places became very important, scientifically and personally. In Sydney, collaborations with Tim Bedding, Dennis Stello, and Daniel Huber were central to many important improvements of the project and the science, and I enjoy working with those people and the groups. In Birmingham, it was a pleasure to work with Bill Chaplin and his group, and they played a major role in solar-like oscillations and in the link to the *Kepler* exoplanet programme. In Porto, Margarida Cunha, Tiago Campante, Mário Monteiro, and others helped build a very strong asteroseismic environment that has become an important part of both KASC and TASC. At MIT, the connection to the TESS mission, including George Ricker and Roland Vanderspek, was important when the experience from *Kepler* was carried into the TESS era. I have also very much enjoyed collaborations with individuals such as Saskia Hekker and many others who helped shape the field. And in Denmark, I worked with Vichi Antoci, Mikkel Nørup Lund, Mia Sloth Lundkvist, Frank Grundahl, Søren Frandsen, Karsten Brogaard, Torben Arentoft, Víctor Aguirre Børsen-Koch, Simon Albrecht, and many students over the years.

These collaborations also led to activities beyond the main *Kepler* and TESS data analysis. One example is the collaboration with the Vilnius group in Lithuania, including Gražina Tautvaišienė, Erika Pakštienė, Edita Stonkutė, Arnas Drazdauskas, and Šarūnas Mikolaitis. They took an active part in observations supporting *Kepler*, TESS, and PLATO, using their telescopes at the Molėtai Astronomical Observatory. For me, this is also part of the KASC and TASC legacy. The consortia helped create networks of people, training activities, observing programmes, and long-term scientific friendships.

The link to the *Kepler* exoplanet mission remained important throughout. Asteroseismology helped determine stellar parameters for key targets, and those stellar parameters improved the planet parameters. *Kepler* was built to find planets, but planets are understood through their stars. In that sense, KASC helped connect the exoplanet revolution to stellar astrophysics in a very direct way.

Many people deserve credit for making this possible. Jørgen Christensen-Dalsgaard provided the leadership and scientific vision that brought Aarhus into *Kepler*. Ron Gilliland and Tim Brown were crucial links between the *Kepler* mission and the asteroseismic community. Bill Borucki, Dave Koch, Natalie Batalha, and many others in the *Kepler* project gave KASC the opportunity to exist within the mission framework. In Aarhus, Rasmus Handberg deserves special thanks for the development of KASOC, with important support from Søren Frandsen, Danijela Jelicic, Brigitte Henderson, Louise Børsen-Koch, and many others. I am also very grateful to the working-group chairs and members, who turned the structure into science.

The later creation of TASC, thanks to George Ricker, David Latham, Sara Seager, and Roland Vanderspek, built directly on this experience. TESS had open data from the beginning, so the situation was different. Even with open data, a scientific community benefits from structures that help people find each other, avoid unnecessary duplication, and develop projects together.

I wish to emphasize a very important part of KASC and TASC: the personal engagement of Bill Borucki and George Ricker. In both *Kepler* and TESS, they understood from the beginning that asteroseismology could not simply be added as an additional science case. It also had technical requirements that had to be built into the observation programmes. As PI of the *Kepler* mission, Bill saw very early the potential of *Kepler* asteroseismology and worked to make sure that the short-cadence resources not needed for the exoplanet programme could be used extensively for stellar oscillations and variability in general. He also supported the efforts to observe stars that were very different from the typical *Kepler* planet-search targets, including saturated stars. Here, the work and experience of Ron Gilliland and Steve Bryson were essential, allowing us to obtain excellent data for many stars for which data would otherwise have been unavailable. This was especially true for the brightest and very saturated stars needing custom-made masks, which Steve Bryson worked out how to integrate into the programme.

As PI of the TESS mission, George Ricker played a similar role for TESS. From the start, he wanted a strong asteroseismic component in the mission, invited the community into the process, and took an active part in TASC workshops and discussions with scientists from many parts of the consortium. His support was also crucial in making the 20-second observing mode operational sufficiently early that it could be used for asteroseismology, once approved later in the mission. This opened a new and very exciting opportunity for studying high-frequency oscillations in bright stars across almost the whole sky. I have had many discussions with both Bill and George about what might be possible if the missions could be suitably adapted. I was impressed and delighted by their personal commitment and by their interest in understanding the possibilities, needs, and technical requirements of the asteroseismic community. KASC and TASC became central parts of *Kepler* and TESS not only because the science was strong, but also because Bill and George worked actively to make the necessary technical capabilities, target-selection policies, procedures, and scientific opportunities happen.

My advice to future consortia would be to keep them as open as possible, but not structureless. A good consortium needs bottom-up initiative and enough top-down coordination to make the collaboration fair and effective. It should encourage new ideas, protect early-career scientists, make responsibilities clear, and give credit to the people who do the work. It should also create places where people meet regularly, because many of the best ideas and collaborations start in discussions over coffee, dinner, or a workshop session, not only in formal documents. If that balance is achieved, a consortium can become the place where a field develops.

For me, KASC was such a place. It helped turn *Kepler* asteroseismology from an opportunity into a community effort, and it helped make asteroseismology part of the wider framework of modern astrophysics. It was a privilege, and a great pleasure, to be part of that process.

## Ronald (Ron) Gilliland

*Emeritus Astronomer, STScI, USA*

### From possibility to formation

This was covered well in my 2014 paper "Prelude to, and Nature of the Space Photometry Revolution" [1], although that paper was probably more related to the needs in the scientific community than the initial idea behind forming KASC. It would be fair to say that KASC might well not have been formed without my and Tim Brown's participation as representatives of the *Kepler* Science Team with interests in both exoplanets and asteroseismology. It would be even more fair to say that KASC would not have been formed without the primary ideas of Jørgen Christensen-Dalsgaard and Hans Kjeldsen.

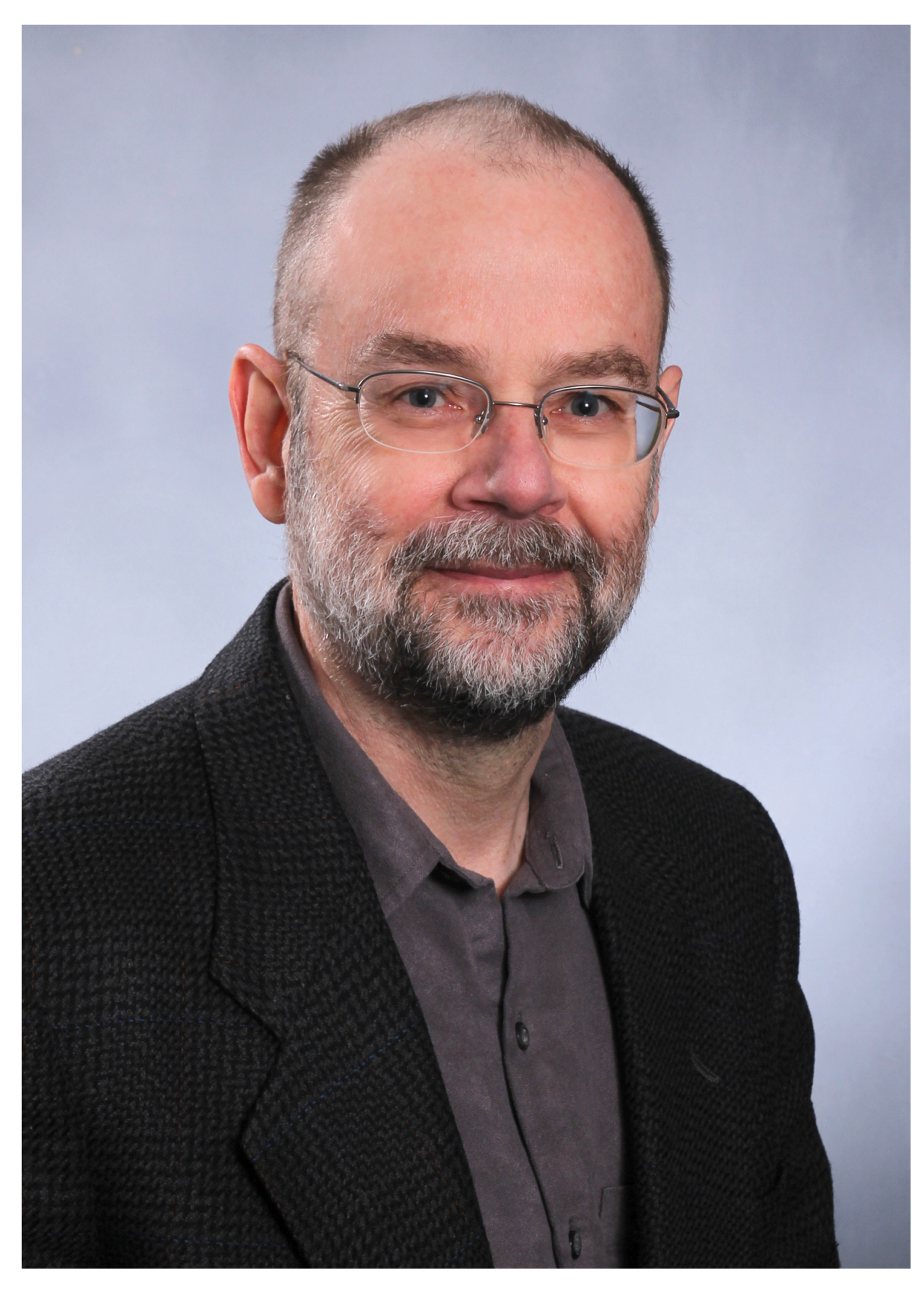

The community had arrived at a point around 2004 where it looked like the *Kepler* Mission would become a reality, and likely provide revolutionary data for asteroseismology. At that time, monetary support for making use of the resulting asteroseismology data was completely dropped. That was, of course, not something I was happy about, but I was thrilled that we (Tim and I) were able to convince the powers that be (*Kepler* PI Bill Borucki primarily) that the capability to obtain short-cadence asteroseismology data should be preserved even if funding to analyze it with this NASA mission had been dropped.

### The founding circle

I would consider myself, Tim Brown, Jørgen and Hans to be the founding members of

[1] `https://doi.org/10.1051/epjconf/201510100001`

KASC. The early formal documents tended to be between myself, Jørgen and the *Kepler* PI, William Borucki.

From my perspective, Tim Brown and I had been pursuing asteroseismology from the ground together for about 15 years before the consortium was formed. Jørgen and Hans had an equally long collaboration attempting to obtain asteroseismology from the ground. Both groups as well were active in trying to promote space missions that would be better suited to delivering broad ranging asteroseismology results. We had some minor collaborations between the groups, for example we were all part of the ground-based network of 4-m telescope observations of M67 I organized in the early 90s for asteroseismology. Myself, Tim and Hans were the primary investigators. If we had been so fortunate as to solidly detect oscillations then I am sure Jørgen would have been deeply involved for interpretations.

### Motivation and momentum

Briefly put, Tim Brown and I had succeeded in getting the *Kepler* Mission to include asteroseismology capabilities and data production, but failed in securing funding for analyses. In a symmetric way, Jørgen and Hans had failed to maintain association with a major planned mission, *Eddington*, but had secured national support for analyses. Given our respect for both Jørgen and Hans, it was natural to reach out to them and attempt to broaden collaborations in order to maintain some capability of our participating in real results from the *Kepler* Mission.

But, personally, I would not have formed a consortium such as KASC became. I view Hans and Jørgen as being the driving forces behind that. It was clear to Tim and me that even if we had gotten good funding support the full range of asteroseismology promised from *Kepler* would be beyond our ability to cover. This remained true even if Hans and Jørgen became deeply involved, hence the impetus to form a much broader consortium was well founded.

### In the middle: Trust, filters, and friction

The first two words in KASC are *Kepler* and Asteroseismology. To include asteroseismology within the *Kepler* Mission made perfect sense from any practical considerations. The returns from asteroseismology were expected to be able to support exoplanet interpretations, the primary mission goal, while maintaining the capability on the spacecraft for asteroseismology was both low cost and low risk. However, the broad, international consortium proposed as KASC was viewed with some considerable lack of support and suspicion by the mission PI, and perhaps NASA.

One example of a practical aspect that proposing KASC meant, and the associated institutional mistrust was a proposal from the mission that before any asteroseismology data could be shared with KASC it would have to be run through a filter that would destroy any signature of transits while (hopefully) maintaining full asteroseismology information. I attempted to argue vigorously that we did not need to and should not require application of this transit-destroying filter to the data before sharing with Hans, Jørgen and KASC. In parallel Hans, and members of the core *Kepler* team volunteered that they could easily make such a filter. As the primary go between from the *Kepler* Mission to KASC all of the data from the primary mission that would be seen by KASC for the first two years went directly through me. I would be the one running the transit destroying filter on all of the short-cadence data before sending it off to KASC. This was not an enjoyable task in any way for me! I resented having to spend 2-3 full days every three months applying the transit-destroying filter, and shuffling data for something that I viewed as basically stupid to do. It felt to me that I was put in this situation of having to degrade the asteroseismology data through the combined lack of good judgment by both the mission (the PI, Bill Borucki), and KASC (Hans in this case). The transit-destroying filter, and other similar aspects of being in the middle between the mission and KASC surely led to a decline in relations between myself and the Aarhus contingent of the original KASC. That (decline in relations) remains an unfortunate outcome of challenges faced in the early stages.

The 15 years supporting proposals to have *Kepler* chosen, then work to support its capabilities being good for both asteroseismology and exoplanets remain probably the greatest highlight of my career. I am proud of a small, but probably very important role in facilitating asteroseismology from *Kepler* to become such an obvious success. The couple of months after launch when I was the only person allowed access to *Kepler* asteroseismic data, and being able to see obvious asteroseismic signals in low-luminosity giants, subgiants and many dwarf stars remain the most fun science period of my life. Unfortunately the challenges of being put in the middle between the mission and KASC in terms of data delivery in the months and couple of years thereafter became a very unpleasant time for me.

### Revolution and reflection

The most significant impact was simply that *Kepler* in particular was able to revolutionize the field of asteroseismology. That happened within the KASC consortium. An interesting question, well beyond my ability to provide an answer, would be whether asteroseismology would have developed well from *Kepler* data being made public without KASC. A valuable aspect of KASC came through providing diverse input from a broad range of experts in guiding what observations would be obtained with *Kepler*.

It would be fair to say that although a core founding member of KASC I never viewed myself as part of the consortium. I viewed myself as a member of the core *Kepler* Mission science team for which the creation of KASC became the way to make use of the mission observations for asteroseismology. It was the best of times and the worst of times for me. My having first access to the data allowed me to see that the efforts expended in earlier years were going to pay off in wonderful ways in terms of both the quality of the data, and the science almost surely to follow from that. Some of the collaborations within KASC that became the most positive memorable experiences involved Bill Chaplin, who I do not recall having any association with before KASC, and Tim Bedding with whom I had corresponded a fair amount. The first results on dwarfs led by Bill, and first on giants led by Tim were immensely rewarding to be a part of. The day-to-day drudgery of being the go between with the mission and KASC regarding data delivery and interactions became, and remain memorably poor experiences.

## Conny Aerts

*Professor, KU Leuven, Belgium*

### From CoRoT to KASC

I was heavily involved in the CoRoT space mission, and we had been holding biannual CoRoT weeks to prepare the optimal target selection and exploitation of its data. The satellite pointing required compromises between the exoplanet and asteroseismology communities, and it was quite enlightening (and challenging) as an astrosociology experiment! It was obvious to me that such interactions to get the most out of a space mission doing white-light photometry make the community optimally prepared. So the need for a consortium was inspired by the CoRoT case for me, although the field of view of the *Kepler* mission made things a lot easier: the exoplanet science case determined the mission in terms of pointing, so no discussions on that were needed.

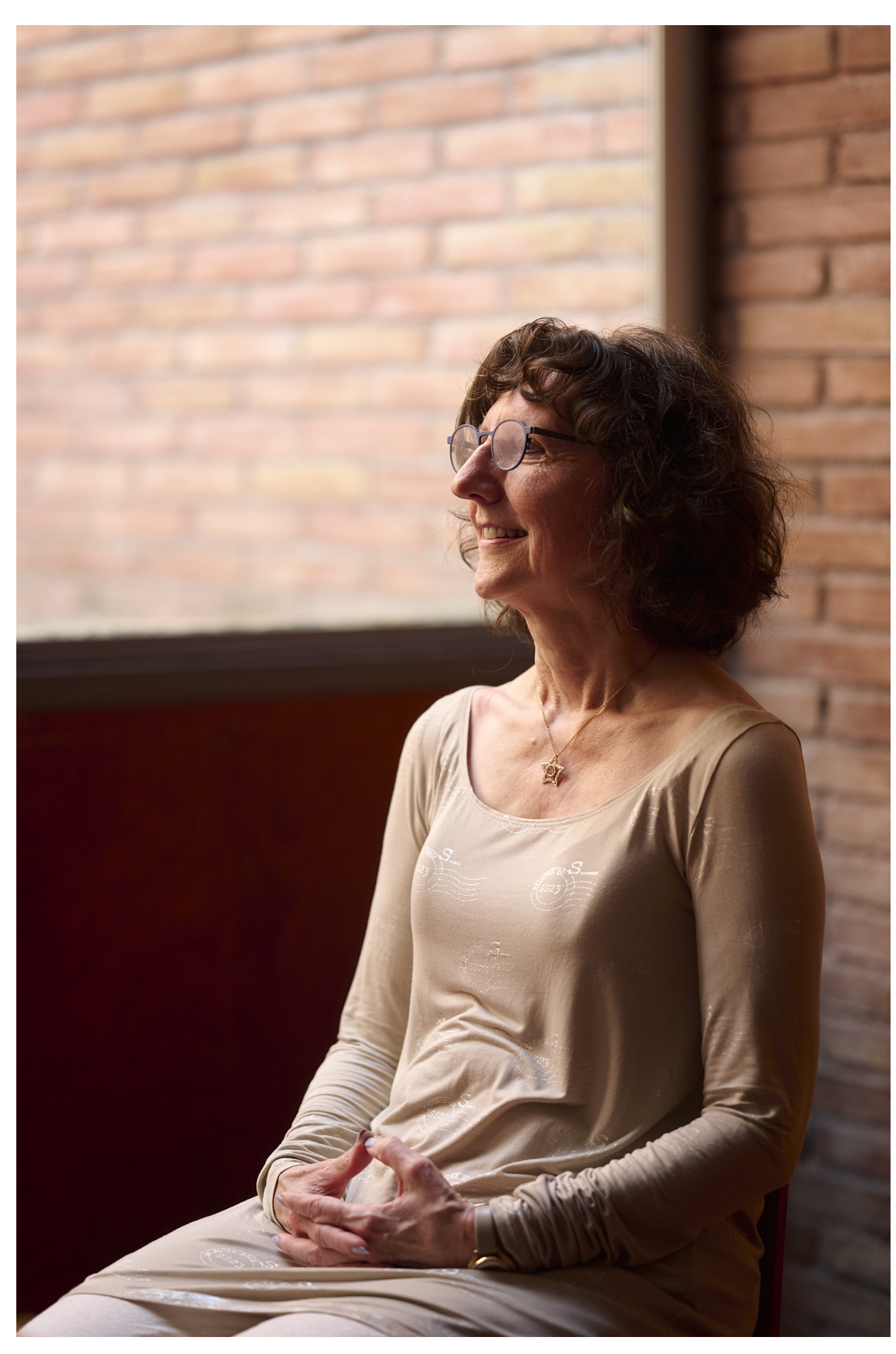

The CoRoT data analysis showed that being prepared by understanding the noise sources to be expected from the mission gives a great advantage once the data start coming in. It also showed that organising the satellite community in working groups (on the type of pulsators to study) was working well. And, of course, we were all very eager to work on the *Kepler* data, so I happily accepted to be involved in the Steering Committee created to define the policies for and guide the practicalities of the KASC consortium.

The global plan was to get organised in terms of working groups, each specialised in particular types of stars. As it turned out, there was the need to have a special working group on pulsating binaries that was a little more complex than the other groups, because the *Kepler* team itself had an eclipsing binary group to help the planet hunting and modelling of the eclipses (since this is so similar to planetary transit detections and modelling). Given my expertise in organising the community in rather "complex" situations (cf. my CoRoT experience) and my good collaborative spirit with respect to the *Kepler* EB WG led by Andrej Prša at Villanova, I volunteered to lead the KASC working group on pulsating binaries.

### Anticipation and scientific motivation

We started holding yearly KASC meetings, much in the spirit of the CoRoT weeks, only far simpler, as we did not need to discuss the selection of the field of view. Most of the discussions by the "sun-like" WGs went into the filter that Hans Kjeldsen had to design to prevent us from discovering exoplanets. But I was not interested in that: anyone with a little bit of knowledge on statistics and missing data knows that the filter "secret" would be easy to unravel… so I was focusing on something much more interesting (see below).

My Leuven team was also still highly focused on the CoRoT data of B-type pulsators in the time frame 2006–2010. And also on ground-based multisite campaigns of Beta Cephei stars to hunt for their internal envelope rotation, because CoRoT had observed only one such star that turned out to have only a dominant non-linear radial mode, aside from some stochastic oscillation signals of unknown origin.

A breakthrough was reached when the CoRoT 5-month-long light curves led us to the first detection of a period spacing pattern of high-order gravity modes in a Slowly Pulsating B-type star (Degroote et al. 2010, published in Nature). This was a slow rotator, and we wanted to get an understanding of similar observational diagnostics across the entire range of possible rotation rates, from slow to almost critical. This was a major motivation for us to join the KASC consortium. However, nobody knew how the period spacing pattern of a fast rotator would look, let alone how long the data strings would have to be to detect it…

### Taking a strategic detour

Doing asteroseismology of stars more massive than the Sun needed patience… When the first *Kepler* data strings came in, we in Leuven shared our time between analysing more CoRoT data while preparing for the years-long *Kepler* light curves of massive stars. We knew from CoRoT that it would take at least one year to find breakthroughs for the internal rotation of massive stars (B-type and F-type dwarfs), so I decided to first try something else: hunting for mixed modes in red giants, following the theoretical predictions by Marc-Antoine Dupret et al. (2009)—again triggered by our Belgian involvement in CoRoT. The CoRoT exoplanet data led to the observational detection of non-radial oscillations in red giants by my colleague Joris De Ridder et al. (2009, also a Nature publication)—luckily, the CoRoT exoplanet experts had failed to avoid having numerous red giants among the selected stars. We knew from theory that mixed modes are essentially also gravity modes in the deep interior of the star and, given that they act in slow rotators and have periods of only a few hours, we knew it would be easier to find period spacing patterns for these "sun-like" pulsators should they reveal mixed modes. Once found—independently by teams in Leuven, Sydney, and Meudon—detecting the rotational splitting of mixed modes was only a matter of waiting until the *Kepler* light curves became long enough. If accessible, it would give a direct view of the internal rotation. It turned out that the rotational splitting popped up once the *Kepler* strings were 2 years long.

### Impact and legacy

For us in Leuven, it was a memorable time when we started to understand the major effect of the Coriolis acceleration on the period spacing patterns of high-order gravity modes in both SPB and Gamma Doradus pulsators. We discovered that the patterns of prograde dipole modes reveal a "tilt" that is directly connected to the rotation frequency adjacent to the convective core of these stars. It was thanks to our CoRoT expertise that we were able to "see" the patterns in the data and to understand what they meant. This beginning of gravito-inertial mode asteroseismology—it took us years to develop it—is, I believe, the biggest achievement of the Leuven team in the field of space asteroseismology.

In my biased opinion, the theory of the internal rotation and angular momentum transport inside stars across the entire stellar evolutionary paths is the most important asteroseismology success. It could not have been unravelled with any other method than asteroseismology. We, as a consortium, made this an observational science with a large impact. The models of stellar structure and evolution were up to a factor 100 wrong in their assumptions of the internal rotation frequency of stars. This has major implications for internal mixing, for age-dating, and for chemical yield computations. The (extra)galactic science communities have yet to absorb the new information…

### Memorable experiences

Well, I am of course biased, but I have always considered the 2022 Kavli Prize in Astrophysics and the 2024 Crafoord Prize in Astronomy as recognition of the field of asteroseismology. In that sense, they are "community awards". These prizes are at a level of the highest recognition. It has been a great honour to have been picked (twice) as representative of the asteroseismology community, with thanks to Hans Kjeldsen for coming up with the nickname of "Queen of Asteroseismology" after a hilarious encounter in Porto (after jogging along the Douro river during a holiday for me, work trip for Hans). The prize weeks in Oslo and in Lund–Stockholm have been the most memorable weeks in my scientific career.

### Outlook

New "large survey" instrument projects for astrophysics are ongoing or on the horizon, from space and from the ground: SDSS-V, WEAVE, 4MOST, PLATO, Rubin, Roman, ET2.0, and hopefully HAYDN. While PLATO, ET2.0, and HAYDN are specifically designed for asteroseismology from uninterrupted space photometry with long duration (years) and high precision (parts-per-million), the other surveys, including Gaia's Data Releases 4 and 5, will also reveal capacity to add to the field in a way to treat large populations in one homogeneous way. We are thus heading towards the standard application of asteroseismology at galactic scale for the Milky Way, the LMC and SMC, and perhaps other galaxies in the Local Group.

On the lessons learned: Keep any consortium fully open, with bottom-up collaborative spirit and while maximising diversity in the broadest sense. Be more proactive in reaching out asteroseismic results towards other communities outside stellar astrophysics. Encourage and embrace scientists from other fields who would like to become acquainted with asteroseismology and help them to get the expertise and skills to actually apply asteroseismic modelling.

## Donald (Don) Kurtz

*Professor Emeritus, University of Lancashire, UK*

In the early 1980s, I was working at the South African Astronomical Observatory (SAAO), making some of the most precise stellar photometric measurements at noise levels in amplitude spectra of a few tenths of a mmag. This led to an invitation to be one of 20 participants at the "Workshop on Improvements to Photometry", held 18–19 June 1984 at San Diego State University (SDSU). The workshop was organised by W. J. "Bill" Borucki of NASA Ames Research Center and Andrew "Andy" Young of SDSU.

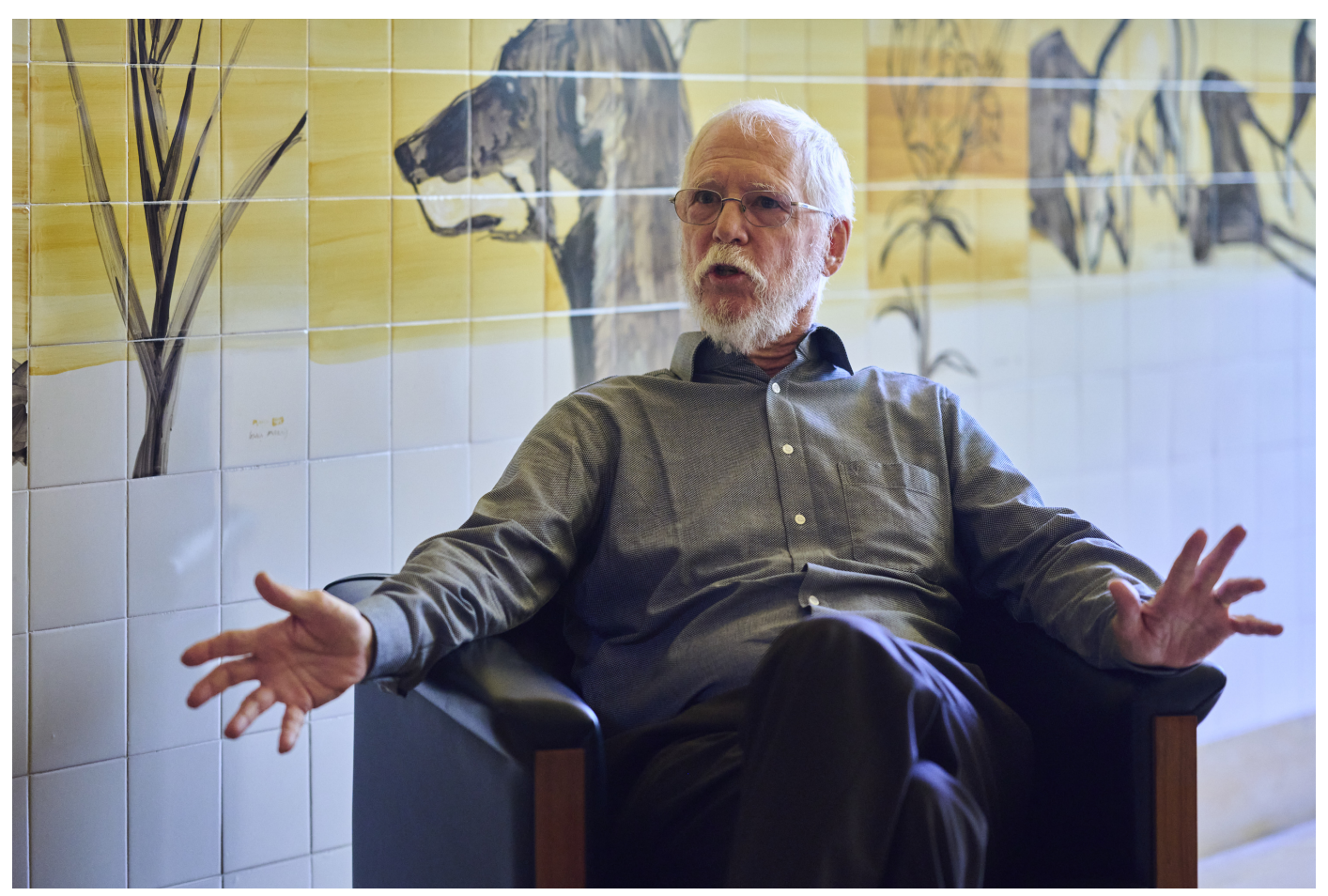

"Five major categories of problems that could be addressed by high-precision photometry were listed: (1) ordinary stellar variability at low amplitudes including spottedness; (2) asteroseismology; (3) extragalactic variables, particularly galactic nuclei; (4) transients such as supernovae, novae, optical counterparts of gamma-bursters, planetary transits, and flares; and (5) solar-system objects such as the zodiacal light, coronal plasma clouds, comets, and asteroids." [2] Technical talks discussed instrumental methods to improve ground-based photometric precision, and the participants "enthusiastically anticipated" another meeting in a year or two.

Of course, the main topic of interest for Bill Borucki was the search for Earth-like exoplanets by the transit method. On the evening of the second day, after the workshop concluded, a conference dinner was hosted by Bill at a fine Mexican restaurant in a hacienda in Old Town, San Diego. We indulged in such large margaritas that I was in no condition to drive after the dinner, so after the others had left, I sat out on the porch of the hacienda and mulled over what we had learned. I came to my own conclusion that no Earth-like planets would be detected by transits observed from the ground, no matter how much we improved the detectors and our techniques. And a space telescope dedicated to photometry seemed like pie-in-the-sky—no prospects for that, I thought.

[2] From Ames Research Center Proc. of the Workshop on Improvements to Photometry p 2 (1984).

I did not know Bill Borucki; his drive and perseverance. Twenty-five years later, the *Kepler* Mission was launched and it revolutionised stellar photometry. From the ground, I could work on one star for a season—months—measuring its brightness variations all night with high-speed photometry and regularly reaching a precision of 0.1 mmag—a part in $10^4$. But I only had the telescope for one week a month. There were gaps in the data because of the day-night cycle; because of weather; because of competitive telescope schedules; because the Earth orbits the Sun and targets disappear into the daylight hours.

Then *Kepler* gave us a 4-year data set for 150,000+ stars with essentially no gaps in the data set at a precision of a part in $10^6$! The precision was 100 times better than for ground-based observations, and the usefulness of the data with nearly gap-free light curves multiplied that improvement of 100 times immeasurably more. My high-precision work at SAAO for more than two decades became obsolete, and I could not have been more delighted.

Almost everything we looked at led to discovery. There is a general rule in experimental science: I call it the *Tychonic Principle* after Tycho Brahe's passionate drive for high-precision observations. If measurements are improved by orders of magnitude—if you can "see" 100 times better—you are going to make discoveries, and many of them come as surprises. This is the same idea that drives particle physicists to build ever more powerful accelerators: you do not know what lurks over the horizon.

While *Kepler* examined a small field for 4 years, TESS has now given us all-sky photometry with some objects in the continuous viewing zone with 1-year data sets. In asteroseismology, and more generally stellar photometry, we used to have WET—the Whole Earth Telescope. The consortium of $\sim$100 astronomers and many observatories was dominant in obtaining extended data sets of a few weeks with the best duty cycles then possible. It was a vast amount of work and organisation. I look back at WET data sets for old stellar friends and then, at a few clicks on my computer, download vastly superior TESS data for a fresh look and usually better understanding for those stars.

Of course, the scientific highlights for both *Kepler* and TESS have been the astounding success in the discovery of exoplanets. Arguably, the observational and theoretical study of exoplanets, and the search for extra-terrestrial life, is now the most dominant field in all of astronomy, even eclipsing the previous leader, cosmology. *Kepler* was not planned and built for stellar astronomy. But it was quickly evident that to understand the exoplanets, you must understand the host stars. TESS was designed with that in mind. Revolutionary highlights in stellar astronomy from these two missions include:

- Astrophysical inference for solar-like pulsators in their thousands (up from just the Sun and a few others) and for red giants in their 10s of thousands (up from the several hundred found in pioneering discoveries with data from the CoRoT mission).
- The expansive understanding of stellar rotation from gyrochronology of cool stars.

- Internal rotation and core-to-surface rotation, internal angular momentum distribution and transfer, with impact on stellar evolution models, hence isochrones, and understanding the ages of stars, clusters, galaxies and the universe.
- High-accuracy determination of the radius, mass and age of stars with critical impact on exoplanet characterisation, and for age on Galactic archaeology.
- Asteroseismology of upper main-sequence stars: Beta Cep, SPB, delta Sct, roAp and gamma Dor stars. We did not understand that we needed long continuous data sets to resolve the g modes in SPB and gamma Dor stars until we got the 4-year *Kepler* data.
- Binary-star light curves in their thousands. These are the bedrock of stellar fundamental parameter determinations with an impact on all of astronomy.
- Tidal interactions on pulsations and vice versa in close binary stars: heartbeat stars, tidally-tilted pulsations, triaxial pulsations, apsidal motion, circularisation.
- Asteroseismology of compact and evolved stars: white dwarfs and subdwarf stars.
- Deeper understanding of the all-important distance indicator stars, the RR Lyr and Cepheid variables.

The *Kepler* and TESS missions have revolutionised stellar astronomy. How did the community come together to work on this? The story of how KASC (and then TASC) was conceived is best told by Jørgen Christensen-Dalsgaard, who was Chair of the first Steering Committee. I received an email from him on 6 September 2007 inviting me to join that committee, which I enthusiastically agreed to do. The KASC Steering Committee set up rules that were welcoming and inclusive for anyone who was interested to join the consortium. Access to the data was through KASC initially, hence there was a strong incentive to join. At first, NASA Ames put a filter on the data to keep us stellar astronomers from discovering "their" planets. It degraded the observations unbearably, and with negotiations was done away with.

KASC grew to many hundreds of members. The most important character of KASC, and now TASC, and their interactions with the exoplanet community, is no doubt collegiality. Science and scientists can be highly competitive, but KASC and TASC were designed to be, and are cooperative. That makes being a part of them a professional pleasure. Lessons learned led TESS to have open access data from the very start of the mission, but our experience with KASC showed that TASC would be useful for collaboration, and so it is.

This does mean that publications are seldom single-author, as they were when I was a young astronomer. In KASC and TASC, we work in groups and consortia. This has many pleasures and strengthens the science produced, as well as the participants' own scientific understanding. But it also needs vigilance to make sure that young astronomers can build their careers, showing their capabilities. Again, I have found this completely cooperative in KASC and TASC. The dominance now at KASC/TASC workshops is obviously from the younger astronomers, with strong participation from both women and men. It is a pleasure to see how vibrant the many fields are that have been spawned by these space telescopes.

Finally, the TASC and KASC websites are very useful. They have been maintained by Aarhus University, but are now in need of further funding for person-power. I would like to see them maintained and continued.

## Sarbani Basu

### *Professor, Yale University, USA*

NASA's *Kepler* mission transformed the field of asteroseismology, evolving it from a niche discipline into a flourishing area of research, attracting many young scientists eager to use and interpret the data. However, *Kepler*'s primary mission was not focused on stellar astrophysics, and there initially was no mechanism to utilize the data for this type of research. This is where KASC (the *Kepler* Asteroseismic Science Consortium) played a crucial role.

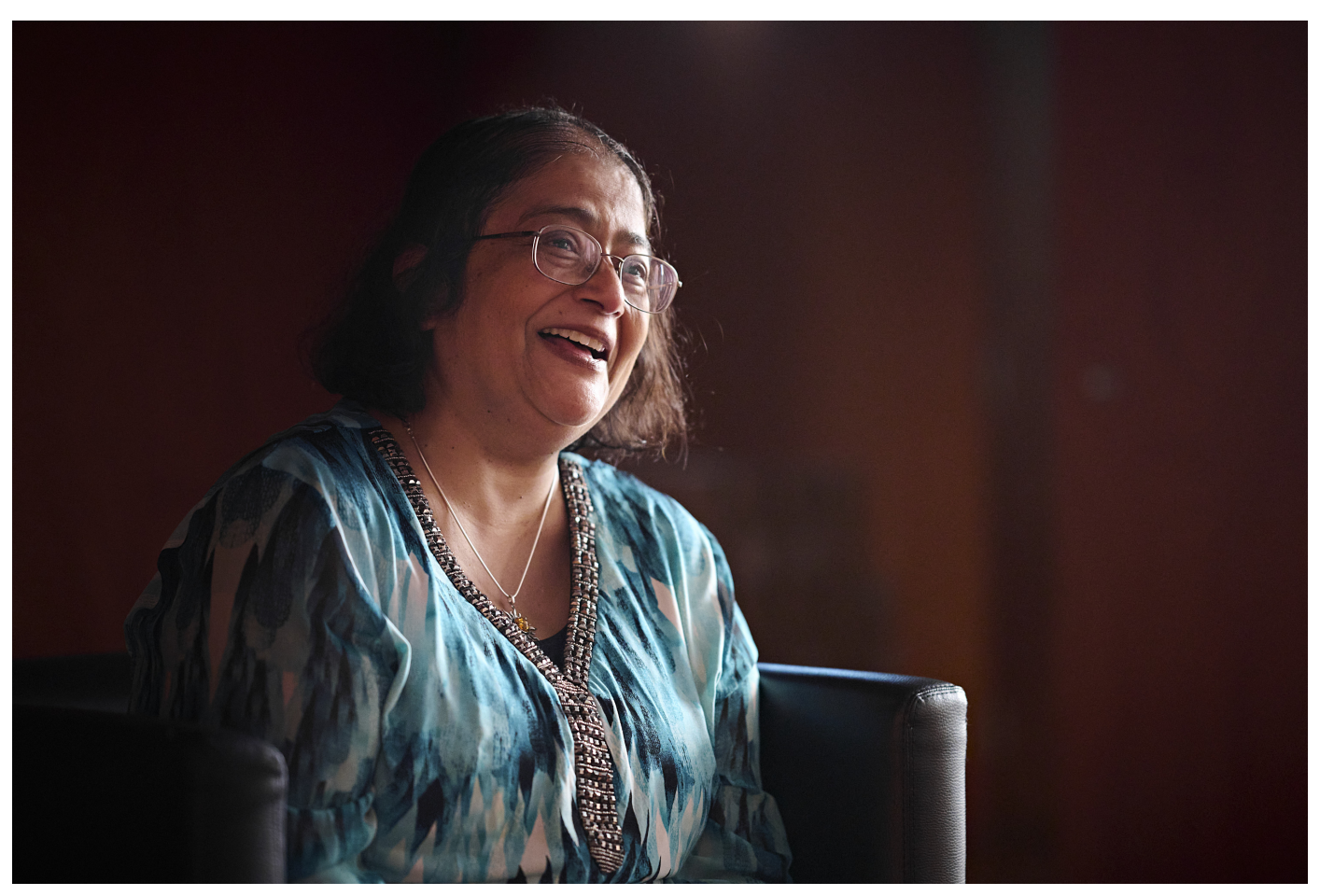

*Kepler* was launched well before NASA implemented an "open data" policy. Consequently, KASC was established to provide data access to asteroseismologists. Beyond simply granting data access, KASC fostered enduring collaborations that continue to thrive today. Asteroseismologists owe much to the efforts of Ron Gilliland and others to convince the *Kepler* team that we were not interested in looking for planets; we just needed the time-series data to study stars using their pulsations. A lot of legwork needed to be done, and that was handled, with aplomb, by Hans Kjeldsen. He first contacted me to be a part of the leadership team and chair the working group on asteroseismology of stars in clusters, usually referred to as Working Group 2, or more simply WG2.

Coordinating KASC in the pre-Zoom era was interesting. We could not easily have telecons where documents could be shared. Most of us did not have easy access to telecon facilities. Fortunately, Aarhus University had a licence for Adobe Connect that could be used; however, bandwidth-related issues meant that we could not really see each other. Very often, in collaborations with the other WG Chairs, we would rely on an old-fashioned phone call. It is quite marvellous how a group of determined people can get together for a discussion, when they have/need to, even with limited technology! Of course, the Earth is not flat, and differing time zones made such virtual meetings difficult (I commiserate with my Australian colleagues; I think they got the worst time of the day to dial in). However, we not only managed, we were very successful.

In some ways, the leadership team was not a "democratically" selected one. But then in those days, very few people had the expertise to work on stellar pulsations (at least solar-type pulsations!), and so all those who did, ended up on the list. The KASC community grew by word-of-mouth; the WG Chairs' collaborators joined, then their students and post-docs. There was a small group of people who refused to join, because they felt that the non-disclosure agreement (NDA) that needed to be signed was against the principles of open science research, and some because the early papers included people who worked on the spacecraft and software pipeline and not the science reported in the papers. However, most of us felt that the NDA was worth signing for data access, and the people who made the data available should be credited. Of course, once *Kepler*'s prime mission ended, NASA made all data public, and the NDA became moot, but many of the collaborations continued. *Kepler*'s successor, TESS, did not and does not have any data restrictions; nevertheless, the TESS Asteroseismic Science Consortium (TASC), modelled on KASC, has been useful in bringing scientists together, and in fostering collaborations.

Of course, one must acknowledge that not everybody was fond of the working group structure because of the policy of WG-wide papers. The problem was that these papers were the consensus views of the authors, and competing views were not readily discussed or were often ignored. I had that reservation too, but I saw the utility of the early papers in popularizing the field, in establishing stellar astrophysics as an important part of science with *Kepler* and in ensuring data access. Of course, competing results were published later, but they had to wait. And given that we continue to find gems in *Kepler* data, I do not believe this is still an issue. The WG structure allowed the community to get some key results published, and that helped the field.

One of the big tasks of WG Chairs was target selection. That was certainly my first task as the Chair—I had to define targets in NGC 6791 and NGC 6819, as well as the much younger (and less populated) NGC 6811 and NGC 6866. I was extremely conscious of the fact that my target selection would affect the science done by many others; until that point, my science was my own, but now I was tasked with enabling science done by others. Since most of the targets were red giants, I also had to coordinate with the Red Giants Working Group (WG8). I recall, the task was quite daunting at first, since we did not have any predictions of mode amplitudes at the time. Another issue was that membership of the clusters was not very well established, despite the best efforts of observers (those were pre-*Gaia* days, after all!). In fact, the first month's data from *Kepler* showed that some of the supposed red-giant members of NGC 6791 could not be so. Nevertheless, my colleagues and I were able to get a good initial target list, which we modified as we learned more from each quarter's data. I just wish there were more, less distant, clusters in the *Kepler* field, so that we could study solar-like pulsations of main-sequence stars, as well as subgiants in the cluster. We did get data on some subgiants in NGC 6819, but their pulsations were, unfortunately, hidden in the noise.

The WGs were essential for the early success of *Kepler* asteroseismic investigations. They brought together researchers who would not normally have worked together. These collaborations led to some very interesting results, such as showing that core-helium burning red giants can be distinguished seismically from inert core ones. The data also showed that we could study the rotation of red-giant cores, thanks to the detection of mixed modes. Ironically, we know the rotation rate of the solar core less well than that of red giants! In fact, the nature of the change of core rotation as a star evolves has become one of the foremost mysteries in stellar astrophysics, since current models of internal angular momentum transfer do not explain the seismic observations. Of course, this is just a small taste of all the inferences that have been (and still are being!) made possible with *Kepler* data.

The early data also allowed us to produce catalogues of stellar global properties, including ages. This revolutionized the field of Galactic Archaeology—with precise ages available, it became possible to infer the history of different Galactic stellar populations. With the availability of velocities and distances from *Gaia*, and ages from *Kepler* and TESS, this field looks very different now from what it used to look like.

KASC made asteroseismology as a technique to study stars visible and popular. There are now more young practitioners of the technique than ever before. In fact, at the 2024 joint TASC/KASC meeting in Porto, Portugal, the participant list was dominated by students and postdocs. This is a very healthy development that bodes well for the future of the field. Within a few years, we will have even more asteroseismic data from PLATO and also some from the *Roman* Space Telescope. Thus, it might be a good time to rethink the structure of the consortium and examine how it can help bring practitioners together, and perhaps, keep coordinating efforts across the community. The KASC and TASC paper repositories have been essential for keeping abreast of the developments in the field, and a specialized repository like that should continue.

I am proud of my association with KASC. Its lasting legacy, in my opinion, is twofold: it enabled the field of asteroseismology to become mainstream, but its bigger contribution is teaching stellar astrophysicists to work together in large groups. We were used to working in small groups, and learning to work in a large collaboration took time. The discipline needed, and the acceptance of hierarchy, albeit a loose one, within the collaboration, was at times difficult for all. But we learned how to work under those conditions, and the sheer volume of significant results shows the success of the consortium.

## Tim Bedding

*Professor, The University of Sydney, Australia*

### The working groups

It is a truth universally acknowledged that one of the major successes of KASC was its working group structure. It is perhaps less well-known that this arose from a similar structure that we set up for a mission that never came to pass. In 1997, the Danish Space Research Institute issued a call for proposals for a small satellite mission. Over the next few years, Hans Kjeldsen led the work on MONS (Measuring Oscillations in Nearby Stars). This was to be a 30-cm space telescope that would carry out photometry on the very brightest asteroseismic targets, one at a time. The mission was selected as the payload on a satellite that would be called *Rømer*. Unfortunately, the program was eventually cancelled, but not before it had established some important legacies. To a large extent, MONS served as a dress rehearsal for *Kepler*. Many of us learned about space photometry, and space missions in general, through our work on MONS.

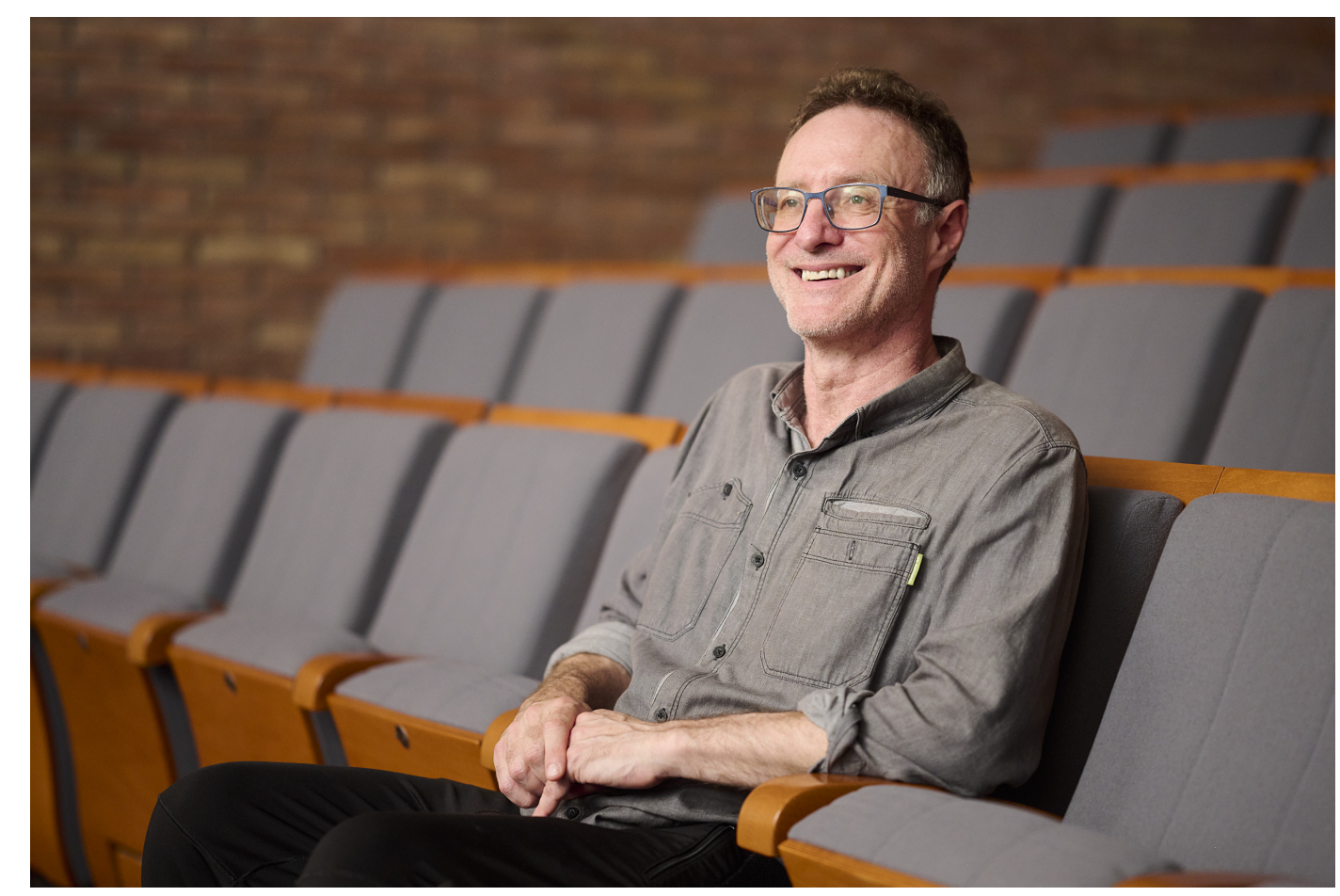

The main goal of MONS was asteroseismology of solar-like oscillations in well-known stars such as $\alpha$ Centauri and Procyon. However, the spacecraft also included a 24-mm wide-field star tracker, which was capable of photometry over a circular field of view 22 degrees in diameter. We realised that this would enable studies of many other classes of variable stars, which is why we set up the structure with working groups on various topics (red giants, delta Scuti stars, roAp stars, rotation, eclipsing binaries, etc.). When *Kepler* came along, it was a natural progression to adopt a similar structure. The KASC working group structure was a major success. Thanks to the great work by the chairs of the groups, it allowed projects to be set up and allocated to researchers. This encouraged a very collegial atmosphere and, importantly, it protected the projects being carried out by early-career researchers. With such a burst of revolutionary data it could easily have been very chaotic. Instead, there was a well-organised and highly productive output that generated a steady stream of outstanding publications.

### 1st April 2010

The formation of KASC allowed us early access to the *Kepler* data, thanks largely to the tremendous efforts of Hans Kjeldsen and Jørgen Christensen-Dalsgaard. However, as the saying goes, there is no such thing as a free lunch—or a free light curve. We were all required to sign non-disclosure agreements (NDAs), confirming that any chance discovery of a transiting exoplanet would be reported to NASA and not made public. Most were happy to go along, since it gave us the chance to play with the most amazing set of data any of us had ever seen. It also gave me a chance to play a memorable trick on Hans and Jørgen.

The scene was the first SONG (Stellar Observations Network Group) workshop in Beijing in 2010. At that time, we were enjoying the flood of light curves that were coming from *Kepler*, but with the full knowledge that any accidental exoplanet discoveries must be treated extremely carefully. The Wednesday of the conference was April 1st, and so I made my plan. I prepared a slide showing my "discovery" of a transiting planet in *Kepler* data. I was extremely excited, explaining that the planet had a period of 3.14 days and the transit depth was exactly what we expect from an Earth-sized planet. (In fact, I actually showed the secondary transit of HAT-P-7b, as published by Borucki et al. 2009. This small dip that *Kepler* had observed when the Jupiter-sized planet went behind its star was impressively small, about the same size as the transit expected from an Earth-sized planet.) After I presented my discovery we broke for lunch, and Hans and Jørgen confronted me. How could I have violated the NDA so flagrantly?! It was only after the lunch break that I revealed the April Fool's joke, and all was forgiven (I hope!).

### The transit removal filter

In addition to the NDA, another condition placed on KASC was that the short-cadence light curves from *Kepler* should be passed through a filter to remove any possible exoplanet transits. The problem was that this filter had the side-effect of introducing artifacts in the data that made asteroseismology quite difficult. Hans Kjeldsen came up with a clever work-around: his suggestion was that, rather than trying to remove any possible transits, they should instead add a few artificial transits to some of the light curves. Knowledge of this would stop anyone using the KASC light curves from trying to publish any exoplanet discoveries, since they could not be sure whether their detection was real or injected. Indeed, one would not actually need to inject any transits at all—it would be enough just to tell people that this had been done! I thought this was a wonderful idea but, sadly, it was not adopted by NASA's *Kepler* team. Instead, we carried on with the filtered data until the filtering process was discontinued about one year into the mission.

### Seventeen workshops and counting

The wonderful successes of asteroseismology with *Kepler*—and now TESS—would not have happened without the huge amount of work done by the KASC leadership, the Aarhus team and the working group chairs. We owe them an enormous debt of gratitude. And the same is true for the organisers of the workshops. I think I am the only person to have attended every KASC/TASC workshop since they began in 2007 (see § III). These meetings are always a great chance to catch up with many fantastic colleagues and to meet new ones, and I look forward to many more.

## William (Bill) Chaplin

*Professor, University of Birmingham, UK*

The asteroseismology community owes a huge thank you and debt of gratitude to Jørgen Christensen-Dalsgaard and Hans Kjeldsen. They had—together with Tim Bedding and colleagues in Aarhus—been working in the 1990s on the development of a proposal for a space-based photometer for asteroseismology called Measuring Oscillations in Nearby Stars (MONS), which would have flown on the Danish *Rømer* satellite. This proposal was to play a key role in the founding of KASC, as I go on to explain below.

Meanwhile, the NASA *Kepler* Mission, led by PI Bill Borucki, was formally approved in 2001 as one of NASA's Discovery Class missions. Borucki and colleagues fully appreciated the potential that asteroseismology could offer in supporting *Kepler*'s prime mission goals, specifically characterizing host stars and hence the planets *Kepler* would discover. The MONS/*Rømer* proposal drew the attention of Borucki et al., and it was Jørgen and Hans whom they approached to discuss how asteroseismology could help. Tim Brown and Ron Gilliland would also play key roles in establishing asteroseismology within *Kepler*.

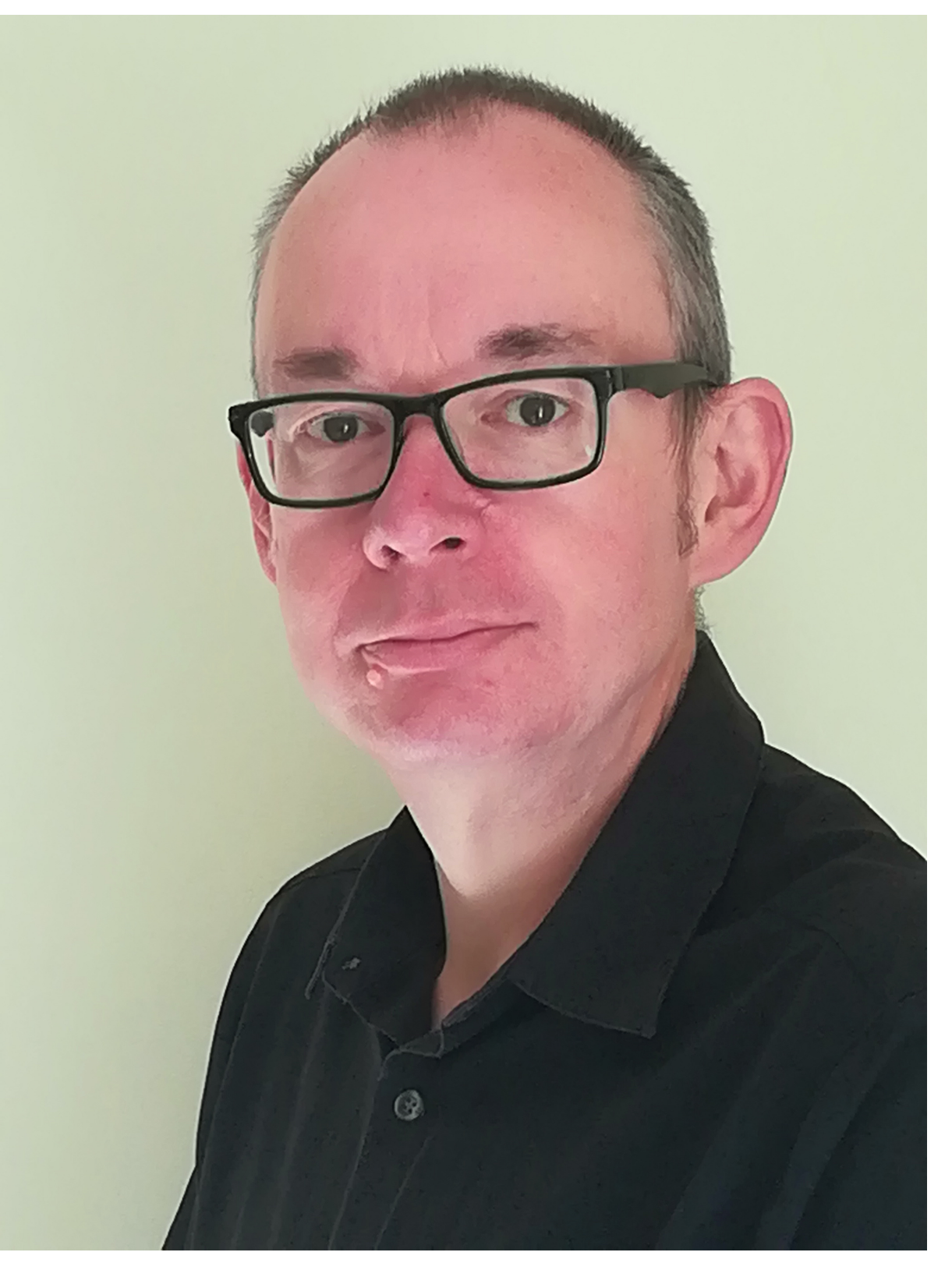

The remit of what was needed implied a significant workload: detecting signatures of solar-like oscillations—ideally in an automated manner—from the analysis of *Kepler* data on thousands of bright targets; when detections were made, extracting global asteroseismic parameters—like the large frequency separation and frequency of maximum oscillations power—and "peak bagging" to extract individual oscillation frequencies; and then using/modelling those extracted asteroseismic parameters to estimate the fundamental stellar properties. In addition to the above, the capability would also be needed to predict whether solar-like oscillations would be detectable in targets, to help inform which main-sequence and subgiant stars should be placed on the requisite (and target-limited) high-cadence list.

From speaking to Hans, I know that he and Jørgen quickly realized that having Aarhus fulfil all these commitments on its own would be a significant challenge; crucially, they anyway held the enlightened view that this was a huge opportunity for the wider asteroseismology community. This was the genesis of the idea to form KASC, and while MONS never became a reality, it had played a vital role in KASC coming into being.

Regarding my involvement in the consortium, the motivation stemmed from the following: I began my research career as a PhD student trying (unsuccessfully) to detect solar-like oscillations in the bright subgiant Procyon A. Whilst my PhD ended up (largely) comprising sets of upper detection limits, this did not deter me from wanting to continue with a research career and for the next few years I pivoted to working on Sun-as-a-star helioseismology, mainly using data from BiSON. Given the ongoing developments in the wider community to detect solar-like oscillations in other stars—notably involving Hans and Tim Bedding—I always felt that at some stage I would want to pivot back to the "dark side". The mid-2000s is when I decided to make that shift of emphasis.

Just prior to then, we had started the solarFLAG collaboration, to bring together members of the international community working on the analysis of helioseismology data. With the strong support and involvement of international collaborators, we started to work together on testing and developing techniques for fitting Sun-as-a-star helioseismology data, from the likes of GOLF and VIRGO/SPM on SOHO, and BiSON. In 2006, with preparations for CoRoT well and truly underway, and *Kepler* nearing launch, we started an asteroseismic version of solarFLAG, called asteroFLAG.

To help get asteroFLAG started, in Spring 2006 we bid into the workshop programme of the International Space Science Institute (ISSI). Our (successful) proposal was focussed mainly on preparing for CoRoT. However, it was the SOHO18/GONG2006/HELAS conference, held in Sheffield in August 2006, that really brought *Kepler* to the fore, placing it at the centre of our asteroFLAG objectives. Hans and Jørgen provided tangible updates on *Kepler* and the potential opportunities for the asteroseismology community (i.e., KASC). In discussions at the conference, it became clear that asteroFLAG had the potential to be an extremely useful framework to help support preparations for *Kepler*, and over the following year we were able to start work in earnest on developing and testing analysis codes that would evolve to become automated codes used for analysing real *Kepler* data.

Jørgen and Hans worked with the *Kepler* leadership, including Tim Brown and Ron Gilliland, to establish the framework under which KASC would operate, overseen by the *Kepler* Asteroseismic Investigation (KAI) Steering Committee, which comprised the above four people. Crucial elements included a formal Letter of Direction with the mission; and a set of KASC policies on data processing and data release—including the application of a transit removal filter to the raw *Kepler* data and the requirement that all members of KASC sign a non-disclosure agreement—together with a detailed publication strategy and rules, with the data and publication policies defined to be consistent with that of the *Kepler* mission, as overseen by the *Kepler* Science Council (KSC). Then there was a proposed structure for how KASC would be organised, with the KAI Steering Committee at the top of the structure, and the *Kepler* Asteroseismic Science Operations Centre (KASOC) acting as a central "hub" and the body responsible for providing asteroseismic results on core mission exoplanet targets (of which more below); there would be a KASC Steering Committee, and individual working groups.

It was during this period that I became Chair of KASC Working Group (WG) 1, "Solar-like p-mode Oscillations", and an inaugural member of the KASC Steering Committee, which in addition to Jørgen, Hans, Tim Brown and Ron Gilliland initially included ten other members. I would also be involved in providing asteroseismic support for the *Kepler* Science Team's core activities (see below). Later I would also become a member of the KAI Steering Committee.

### Early challenges

As was the case for other WGs, early challenges for KASC WG1 included creating a structure for the WG that would facilitate and support the active participation of everyone who wanted to get involved, including the creation of sub-groups and assignment of sub-chairs; establishing clear objectives and science priorities, and a publication plan for the WG; and the creation of a framework and process for proposing targets for short-cadence observation, which were fed into the KASC short-cadence allocation process.

I would also play a role in the asteroseismic analysis that would directly support the *Kepler* Science Team's work on prime mission targets for exoplanet studies. The Letter of Direction formally specified that KASOC would perform this work, with Hans having overall responsibility for KASOC's activities. I was assigned the role as project manager for the Asteroseismic Analysis Chain (AAC) within KASOC, responsible for managing and overseeing all stages of its operation and the delivery of results on *Kepler* Objects of Interest (KOIs) to the *Kepler* team.

I worked very closely with Ron Gilliland on establishing a process for testing whether main-sequence and subgiant targets that had just acquired KOI status would have potentially detectable solar-like oscillations if placed on short-cadence observations. This was an important part of making sure we maximised the benefits that asteroseismology could provide in terms of characterising newly discovered exoplanet systems.

### The impact of the consortium

The first and (I think) most important impact has been the very act of bringing us all together to create a cohesive and collegiate community. We need to remember that before KASC existed, a large international collaboration for the asteroseismology community might have involved a few institutes and perhaps up to 10–15 people. KASC changed this in a marked fashion.

Importantly, KASC also gave the community a coherent identity and voice. Again, we owe a huge debt of gratitude to Jørgen and Hans for recognising what KASC could enable. It was then a natural step for TASC to be created. We must also remember that asteroseismology would arguably not have had the central core-mission role it occupies in PLATO without KASC.

That there are too many science highlights to single out here bears testament to the cutting-edge, world-leading research that KASC's working groups and collaborations have produced. A steady flow of high-impact papers in journals like Science and Nature stands as a fitting legacy; and as KASC led to TASC, that steady stream of cutting-edge publications has continued.

### Memorable experiences

Here, I would put the people involved front-and-centre. Working closely with fantastic scientists from around the world, and getting to know them—not only from a research perspective, but also as friends—has been hugely enjoyable and rewarding. Close collaborations that were initiated in the pre-launch phase of *Kepler* have persisted and flourished. I am sure many colleagues will have similar feelings about KASC and TASC. KASC undoubtedly acted as a catalyst and enabler to bring the community together.

I also hugely enjoyed my regular trips to the Ames Research Centre—NASA's host institute for the *Kepler* Mission—when I was working with the *Kepler* Science Team; I was, for a time, an elected member of the *Kepler* Science Council; and then subsequently a member of the *Kepler* Nominal Mission Closeout Review Panel. It was great to get to know and work with the mission leaders from NASA, and exoplanet colleagues from around the US. I particularly enjoyed working with Ron Gilliland, who I had huge respect for. The meetings were held at the SETI Institute in Mountain View, and I always stayed at the same motel, which felt like it had not been re-decorated for a good 30 years (echoes of the 1980s…) but nevertheless became a familiar reference point on my regular trips.

TASC has created similar long-lasting memories. Arguably, the highlight was attending the launch of TESS in Florida, and spending an extremely enjoyable few days there with Hans Kjeldsen, Rasmus Handberg and Mikkel Lund (where, amongst other activities, we honed our putting skills on a local mini-golf course).

### Legacy, future, and lessons learned

The legacy left by KASC is the central role it has played in asteroseismology having a cohesive community identity. The database of observations available to that community will be further diversified and enriched by upcoming missions and projects, which, of course, includes PLATO. A natural direction for KASC and TASC is therefore one not tied to particular missions, but instead to having a wider identity.

A key foundational element of any successful collaboration is that its members must be incentivised to join and actively participate. Having a clear structure that tensions bottom-up input with top-down strategic priorities that everyone can buy into is crucial. In its early phase, KASC had to work within the constraints set by the data access and usage policies for *Kepler*, but it did so in a way that allowed the asteroseismology community to produce excellent science from very early in the mission, and to therefore take advantage of the huge opportunities that *Kepler* presented.

## George Ricker

*TESS PI, MIT Kavli Institute, USA*

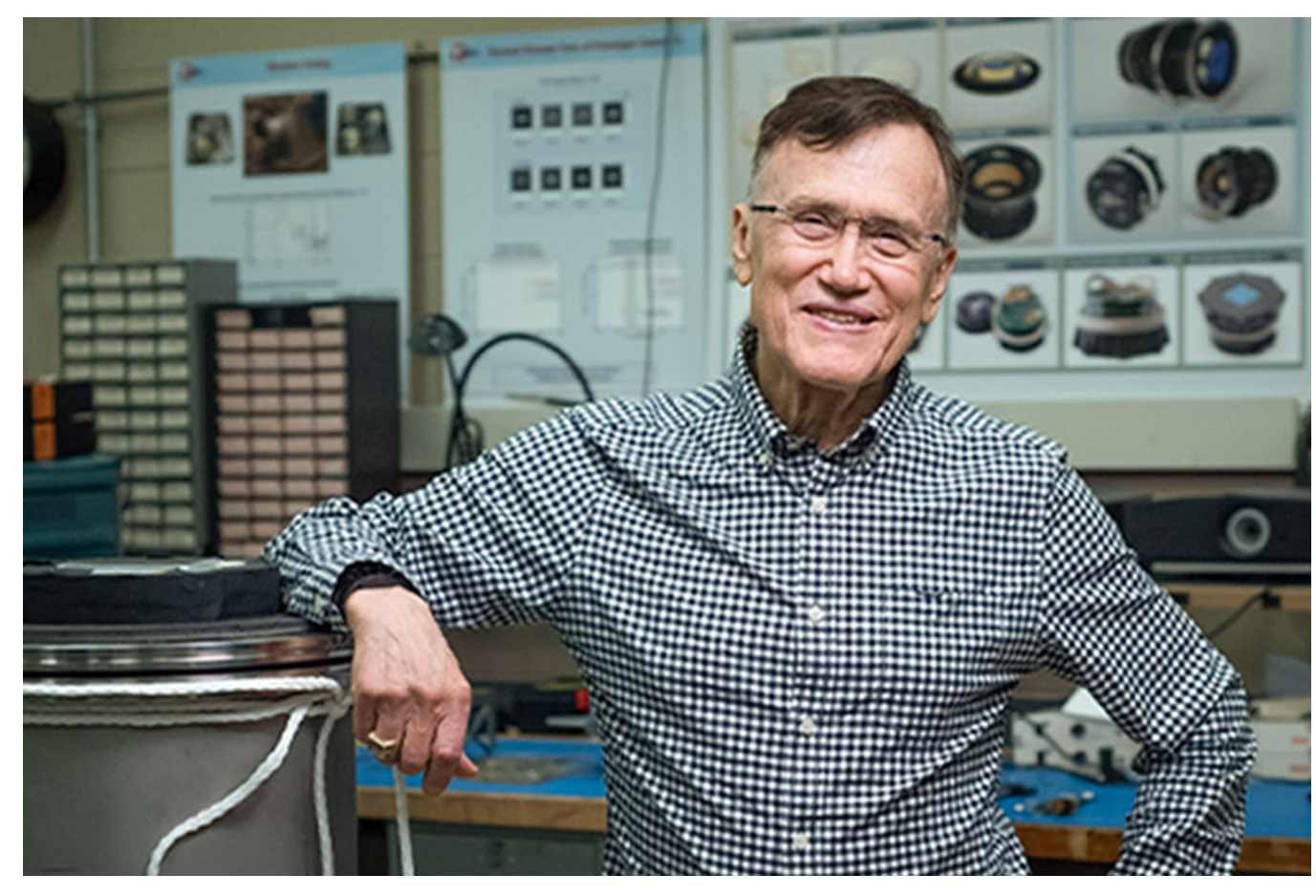

One of the greatest pleasures of my 1.5-decade-long adventure in helping to bring about TESS was witnessing the emergence of a vibrant and welcoming international community centered on asteroseismology and its application to understanding stars and planetary systems. Although my own scientific background was rooted primarily in transient and exoplanet discovery and instrumentation, my introduction to the world of asteroseismology through my early experiences with the *Kepler* Asteroseismic Science Consortium (KASC), and later the TESS Asteroseismic Science Consortium (TASC), became one of the most rewarding aspects of the TESS mission.

When I first encountered the word "asteroseismology", it seemed a bit mysterious and unfamiliar. Yet once I grasped the fundamental concept—that the oscillations of stars reveal their internal structures in much the same way that earthquakes reveal the Earth's interior—I quickly became fascinated with the power and elegance of the concept. I soon found myself sharing my new appreciation of these concepts—and how TESS might contribute to better understanding them—to my colleagues and students on the TESS Team (and at NASA) why this science was not only important in its own right, but also critically relevant to exoplanet research. Precise knowledge of stellar properties ultimately determines the accuracy with which we can characterize planets orbiting those stars. In many cases, asteroseismology provides the most powerful means of determining stellar masses, radii, and ages, thereby enabling more accurate measurements of planetary properties.

My formal introduction to the asteroseismology community occurred well before TESS launched. In April 2013, Sara Seager arranged a visit to MIT by a delegation of asteroseismologists from Denmark and elsewhere in Europe, including Jørgen Christensen-Dalsgaard, Hans Kjeldsen, and several of their colleagues. Their goal was to introduce members of the TESS team to the science and promise of asteroseismology. During two days of presentations and discussions held in the Green Building at MIT, they explained how TESS—following in the footsteps of *Kepler*—could possibly become a transformative instrument for stellar astrophysics while simultaneously advancing exoplanet science. The meeting left a lasting impression on many members of the MIT and Harvard–Smithsonian investigators on TESS, and helped establish relationships that would prove invaluable throughout the mission.

That visit opened the door to an ensuing productive collaboration that continues even more strongly today. Soon afterward, I attended my first KASC meeting in Sydney, Australia, and became immersed in what I often jokingly describe as the "cult" of asteroseismology. What impressed me most was not only the scientific sophistication of the community, but also its collegiality, openness, and sense of shared purpose. The vast majority of experts in the field at that time were located outside the United States, and participation in KASC and later TASC introduced me to a truly international network of collaborators and friends.

The workshop series itself became one of the defining features of that experience. Beginning in 2007, the KASC workshops had already evolved into annual gatherings that were unlike typical scientific conferences. Each meeting felt more like a family reunion than a formal workshop. The atmosphere encouraged openness, collaboration, and the free exchange of ideas among scientists at every career stage. Students, postdoctoral researchers, senior scientists, and mission leaders all interacted as colleagues united by a common scientific purpose.

The first workshop that Roland Vanderspek and I participated in as presenters was TASC1/KASC8 in Aarhus, Denmark, in 2015. That meeting was a revelation. We had the opportunity to present TESS to a community that had already established a rich scientific foundation through *Kepler*. Although TESS was, in some sense, arriving later to a field transformed by *Kepler*, the asteroseismology community welcomed us warmly and immediately recognized the unique opportunities that TESS would provide.

Over the years, these TASC/KASC meetings took us to many memorable locations around the world. Following Aarhus in 2015, we attended the Birmingham meeting in 2017, and then returned to Aarhus in 2018. We then hosted the meeting at MIT in Cambridge in 2019, immediately preceding the first TESS Science Conference. We then traveled to Leuven in 2022, Honolulu in 2023, and then Vienna in 2025. We regrettably missed Porto in 2024 and were unable to attend the 2016 meeting because the TESS team was deeply engaged in critical pre-launch camera testing. Repeated visits to Denmark, Belgium, and other international centers of stellar astrophysics over the past decade strengthened both scientific and personal connections that continue to this day.

One of the most important aspects of the partnership between TESS and TASC has been the consortium's role in shaping observing strategies and mission capabilities. From the beginning, the TESS Project emphasized asteroseismology as a core scientific objective rather than a niche activity. TESS was designed as an open observatory, with immediate public access to data and broad participation by the scientific community. This openness distinguished TESS from many earlier space astronomy missions and ultimately helped establish practices that NASA has since adopted more broadly.

During the development of TESS, the NASA leadership increasingly recognized the scientific benefits of rapid public access to mission data. As TESS Principal Investigator, I strongly supported an open-data philosophy, arguing that immediate community engagement would maximize scientific return. These discussions contributed to a broader shift within NASA astronomy missions toward shorter proprietary periods and more open access to space-based datasets.

The TASC collaboration itself became a model for community organization. Through its working groups, TASC developed an open, collegial, and highly self-organized structure that enabled effective scientific coordination without sacrificing accessibility. The consortium rapidly became one of the dominant scientific users of TESS. Over the first six years of the mission, TASC and KASC members proposed 19,457 of the 32,173 Director's Discretionary Time targets selected for short-cadence observations. Each year I looked forward to receiving extensive target lists assembled by Vichi Antoci and her colleagues, containing fascinating stars whose oscillations revealed unique stellar personalities and scientific opportunities.

The scientific productivity of the consortium has been equally impressive. TASC members have contributed to hundreds of refereed publications spanning stellar structure and evolution, galactic archaeology, binary stars, compact objects, exoplanet host stars, and stellar populations. The influence of TASC extends far beyond traditional asteroseismology, demonstrating how a well-organized scientific community can amplify the impact of a major observatory. Many of the most highly cited TESS papers have involved contributions from TASC researchers, and the consortium has played a major role in establishing TESS as one of the most productive astrophysics missions of its era.

These interactions directly influenced the evolution of the mission itself. Early in the mission, I advocated at NASA Headquarters for enhancements that would especially support asteroseismology, as well as the other core mission goals of transits and transients. The cadence of TESS full-frame images was progressively reduced from 30 minutes at launch to 10 minutes in the first extended mission and then to approximately 200 seconds in subsequent extended missions. These improvements were motivated in large part by the requirements of asteroseismic investigations and substantially expanded the scientific reach of the mission.

Several collaborations that emerged from TASC proved especially rewarding. Among them were interactions with Saul Rappaport at MIT; my student Rahul Jayaraman, whose dissertation was broadened with guidance from Saul from "just extragalactic transients" to also include studies of "tidally tilted pulsators" and "tri-axial" pulsations and new US colleagues who demonstrated how TESS observations could be applied to a wide variety of astrophysical problems beyond the mission's original exoplanet focus. These projects exemplified one of the most important lessons of TESS: when a mission provides high-quality data to an engaged scientific community, investigators from that community will find innovative applications that extend well beyond the mission's original science requirements.

The synergy between exoplanet science and asteroseismology reflects a deeper lesson-learned about the design of TESS. The mission's photometric capabilities were ideally suited to both fields. High-precision, long-duration observations are essential for detecting planetary transits and for measuring stellar oscillations. TESS was designed with an exceptionally large field of view—approximately twenty-two times larger than that of *Kepler*—while simultaneously providing high-cadence observations and unprecedented flexibility in target selection. Its overall étendue, or observational grasp, is roughly five times greater than *Kepler*'s. Combined with the brightness of its target stars, this enabled extensive ground-based follow-up and dramatically expanded the accessible discovery space.

Today, the scientific capabilities of TESS exceed what many of us imagined at launch. Improvements in spacecraft operations, detector calibrations, and data processing have continually increased the quality of the photometric data. New GPU-based processing systems introduced at MIT just this year promise further advances, potentially extending high-quality photometry from the current limit of approximately 13th magnitude to objects as faint as 18th or 19th magnitude. These advances open exciting possibilities for both traditional asteroseismology and emerging applications involving artificial intelligence and machine learning methods applied to variable stars, binary systems, and other complex stellar phenomena.

Perhaps most gratifying, however, has been seeing the impact of TESS on younger generations of scientists. At meetings of the American Astronomical Society and other conferences, I am frequently approached by graduate students, postdoctoral researchers, and even undergraduates who enthusiastically describe discoveries they have made using TESS data. Many comment on how accessible the mission has been and how straightforward the data products are to use. Hearing someone say, "It's all just FITS files", remains one of my favorite compliments to the mission.

As TESS enters its extended future and the astronomical community prepares for missions such as PLATO, Ariel, and the future Habitable Worlds Observatory, the values embodied by TASC remain critically important: openness, collaboration, collegiality, curiosity, and scientific generosity. The partnership between TESS and the asteroseismology community has demonstrated that when new data are shared openly and scientists work together across institutional and national boundaries, extraordinary discoveries become possible.

Looking back over more than a decade of active collaboration with the asteroseismology community, I am grateful not only for the scientific achievements that have emerged from this partnership, but also for the many friendships, mentorships, and sense of community that have accompanied them. The story of TESS and asteroseismology is ultimately not just about stars and planets. It is also about a community of hundreds of dedicated professional astronomers, students, and amateur scientists who have worked together for more than a decade to unlock the mysteries which still surround many of the most basic aspects of the universe in which we live.

# Afterword

The history recounted in this booklet traces the emergence of KASC and TASC as enduring scientific communities. What began as a framework to enable asteroseismic exploitation of *Kepler* data evolved into a model for large-scale, international collaboration in stellar astrophysics. The reflections collected here remind us that the consortia were shaped not only by scientific ambition, but also by negotiation, trust, and a shared willingness to experiment with new ways of working.

In the years since the earliest KASC meetings, the landscape of astronomical research has changed substantially. Data policies have shifted from restricted access and non-disclosure agreements to open archives available to the entire community. The experience gained through KASC informed the creation of TASC, which operated from the outset in a more open-data setting while preserving the collaborative framework that had proven so effective. The field itself has matured: asteroseismology is now routinely integrated into exoplanet characterization, Galactic archaeology, and stellar population studies.

Nearly two decades have passed since KASC's formation in 2007, and equally significant has been the generational transition within the community. Many of the early architects of KASC and TASC now look on as new leaders—often once students within the working groups—that shape the scientific agenda. The workshop series, summer schools, and collaborative structures have ensured continuity while allowing renewal. This intergenerational transfer of expertise and responsibility may stand as one of the consortia's most meaningful achievements.

Looking ahead, forthcoming facilities, including ESA's PLATO, NASA's *Roman*, the Chinese Earth 2.0, potentially ESA's HAYDN, and complementary ground-based surveys, will operate in an era defined by long time baselines and data volume. The technical challenges will evolve, but the central lesson of KASC and TASC remains clear: coordinated, international collaboration accelerates discovery and enhances scientific progress.

If there is a unifying legacy to these consortia, it lies not solely in their scientific output, substantial as it is, but in the culture they fostered: collegial, inclusive, and oriented toward shared progress. As new missions and new communities emerge, that spirit will remain essential.

# Appendices

# A
# Letter of Direction

The following is the signed version of the Letter of Direction, which outlines the structure of the collaboration between the *Kepler* Project and the *Kepler* Asteroseismic Investigation (KAI).

# Letter of Direction

regarding asteroseismic investigation within the Kepler project from the Kepler Project (Project), represented by William Borucki, to the Kepler Asteroseismic Investigation (KAI), represented by Ronald Gilliland.

The present letter directs the activities related to asteroseismology based on data from the Kepler mission. It defines the obligations of the KAI towards the Project, in terms of software development, reporting and delivery of asteroseismic results, as well as the timely provision of the required photometric and other data by the Project to the KAI. In addition, it describes: (1) the Kepler Asteroseismic Science Operations Center (KASOC) set up to manage these investigations, and (2) the Kepler Asteroseismic Science Consortium (KASC), a group of collaborating scientists and/or institutions established to accomplish the activities of the KAI. The letter shall remain in force through the end of Phase E of the Kepler mission.

## 1 Introduction

The Kepler photometric data will represent a unique resource for asteroseismic investigations of the global properties and internal structure of a large number of stars having a broad range of different types. In particular, asteroseismic analysis of these data will provide accurate determination of the radius of a large fraction of the stars hosting candidate planetary systems, as determined from the transit analysis of the Kepler data, as well as estimates of the ages of the systems. Through investigation of a broad range of stars the asteroseismic investigation will also substantially improve our understanding of general stellar evolution, and hence strengthen the use of such modelling to further constrain the properties and evolution of the stars and systems investigated in the Kepler extra-solar planet program. The purpose of the Kepler Asteroseismic Investigation is to ensure that full use is made of this potential to benefit the Kepler investigations of extra-solar planetary systems.

## 2 Asteroseismology Products to be Provided to the Project

- Asteroseismic characterization of planet-hosting stars, particularly radius.
- Ability to distinguish cool giants from main-sequence stars through asteroseismic measures, such as estimates of the stellar radius or the shape of the background power spectrum characterizing the granulation time scale.
- Understanding of general stellar properties, including stellar structure modelling, contributing to stellar characterization.

## 3 Data and Support Provided by the Project to the KAI

The asteroseismic investigation will be based on Kepler Mission data from which information about planet transits has been eliminated, through filtering or by other means.

It is assumed that at any given time 512 targets will be observed at a one-minute cadence. It is furthermore assumed that 25 of these targets will be reserved for the guest-observer (GO) program upon its inception. At mission start all short-cadence targets will be selected for asteroseismology. As planet-hosting systems are detected an increasing number of short-cadence targets will be allocated to the study of such systems; in many cases such targets will also be appropriate for, and will be used for, asteroseismic investigations. However, an adequate number of short-cadence targets will be reserved specifically for asteroseismology throughout the mission.

# 4 KAI Activities

## 4.1 Before launch

- Develop and test a "high-pass" filter, or other mechanism, to remove planet information from data to be used for asteroseismic investigation. This will take place in collaboration with the SOC and will be subject to the approval of the PI.
- Develop a pipeline to extract frequencies or frequency properties from observed time-series.
- Develop a pipeline to derive stellar properties from frequencies or frequency properties.
- Make available to the Project either a separate pipeline or other system to estimate stellar radius and/or identify giant stars, from power spectra based on long-cadence timeseries for cool giants. (Alternatively, at the discretion of the Project, use the power spectra provided by the SOC for these analyses.)
- Organize KASC to support these activities.
- Provide the selection of initial targets for asteroseismic program, as well as asteroseismic targets to be observed throughout the mission, for the Kepler Mission Planning.

## 4.2 After launch

- Provide stellar parameters (particularly size) in a timely fashion to the Project; the goal is to provide the results in three months or less after receiving timeseries data.
- Perform asteroseismic analyses on any additional short-cadence targets upon request from the Project.
- Revise the asteroseismic target list as appropriate. In particular, identify cases suitable for asteroseismic analysis from those targets selected for short-cadence planet-transit observations.
- Comparison of the general stellar properties of cool MS stars with those of evolved stars.
- Ensure timely publication of the asteroseismic results.

## 5 Resources provided by the Project

- Provide access to time-series of 512 short-cadence targets chosen for asteroseismic analysis during the initial roll segment.
- Provide guaranteed access to 240 short-cadence targets available for change each quarter, for asteroseismic analyses throughout the mission. Of the 240 targets available for asteroseismology, at least 140 may be selected by the KAI, and up to 100 may be selected by the Project from transit candidate stars meeting brightness and spectral type criteria expected to allow asteroseismic results.
- Selection of 100 or more (as resources permit) long-cadence targets specifically for comparison with MS dwarfs. These can include cool giants to be observed for 4 years, beta Cephei stars, slowly pulsating B stars, and long-period delta Scuti stars.
- Verify and validate the proposed procedure for 'high-pass filtering' before launch.

## 6 Required activities by Co-I Gilliland

- Carry out the 'high-pass filtering' to remove transit information from data to be made available to asteroseismic investigation.
- 'High-pass filtered' data to be provided to KASOC, within one month after the data have been delivered to the DMC.

## 7 Management of the Asteroseismology Effort

The activities of the KAI shall be overseen by a Steering Committee consisting of Kepler Co-I R. L. Gilliland (head), Kepler Co-Is T. M. Brown and J. Christensen-Dalsgaard, and H. Kjeldsen, Department of Physics and Astronomy, University of Aarhus. Matters concerning data rights and publication will be dealt with by the Steering Committee of the KAI, in collaboration with the Project, as required.

To manage the asteroseismic investigation, the Kepler Asteroseismic Science Operations Center shall be established at the Department of Physics and Astronomy, University of Aarhus. The establishment and activities of the KASOC will be covered from Danish funding through the duration of the proposed activities. No funding for non-US investigators is requested from the Project nor from NASA HQ for this investigation.

## 8 Kepler Asteroseismic Science Consortium

To coordinate the many scientists that will participate in the asteroseismic study, the KASC shall be established immediately upon full signature of this letter. Relevant groups shall be invited to submit proposals for their participation in the KAI, including their contributions in the preparatory phase and the proposed use of the Kepler asteroseismic data. Based on these proposals, participants in the KASC shall be selected by the Steering Committee of the KAI, and individual agreements, consistent with the conditions stated in the present letter, shall be established between the KAI and these participants. These shall include a non-disclosure clause prohibiting any release of information about potential planetary transits that may inadvertently be present in the asteroseismic data, as well as any information that might inadvertently identify a potential planet host.

The KASC will be operated on the principle of open access to the Kepler asteroseismic data. A publication policy shall be established by the KAI in collaboration with the Project, consistent with the contributions made by the participants of the KASC to the specific results being published, and with due account being taken of the contributions of the Kepler Team. Papers based on the Kepler asteroseismic data shall be internally refereed before submission to a journal or preprint server.

## 9 Relation to guest-observer, participating-scientist, and data-analysis programs

It is acknowledged that various community support programs may result from NASA Announcement of Opportunities to participate in the Kepler project, and/or obtain additional observations, and/or research support funding for utilization of Kepler data. Successful proposers with research relevant to asteroseismology have the option to apply to be associated with the KAI, as described in this letter.

US participants in the KAI/KASC may receive funding for asteroseismic investigations as allowed for in NASA research solicitations.

For the Kepler Project:

William J. Borucki 2/16/2007

William J. Borucki

For the Kepler Asteroseismic Investigation:

Ronald L. Gilliland 2/12/07

Ronald L. Gilliland

For the Kepler Asteroseismic Science Operation Center:

Jørgen Christensen-Dalsgaard 2/26/2007

Jørgen Christensen-Dalsgaard

# B
# KASC INVITATION LETTER

The following shows the email sent in 2007, on behalf of Jørgen Christensen-Dalsgaard, to invite individuals identified as being interested in asteroseismology to join the KASC.

```
Date: Fri, 1 Jun 2007 21:39:37 +0200
From: Danijela Jelicic <dani@phys.au.dk>
Reply-To: Danijela Jelicic <dani@phys.au.dk>
Subject: KASC invitation
To: X@some.where

Dear X,

With this mail I wish to invite you, and possibly others
in your group, to join the Kepler Asteroseismic Science
Consortium (KASC).

The Kepler mission, scheduled for launch in November
2008, is designed to search for extra-solar planetary
systems, particularly Earth-like planets in the habitable
zone, using the transit technique. This will entail
continuous monitoring over 100,000 stars for at least 4
years, with a 30-minute cadence. In addition, 512 stars
will be observed with a 1-minute cadence. These stars
are obvious targets for asteroseismology, including
observation of solar-like oscillations. Also, the longer
cadence will be appropriate for several types of variable
stars, including hot, massive pulsators and solar-like
oscillations in giant stars. Further detail on the mission
and the planned asteroseismic activities can be found in
a paper by Christensen-Dalsgaard et al. to appear in Comm.
Asteroseismology (http://xxx.lanl.gov/abs/astro-ph/0701323)

An agreement has been reached with the Kepler PI, W. J.
Borucki, to establish the Kepler Asteroseismic Investigation
(KAI) based at the Department of Physics and Astronomy,
University of Aarhus. This agreement is formalized in a
Letter of Direction from Borucki to the KAI, attached to the
present mail, which defines the rights and obligations of
the KAI. The KAI will be supervised by a steering committee
consisting of Ron Gilliland (chair), Tim Brown, Joergen
```

Christensen-Dalsgaard and Hans Kjeldsen. To help preparing for the asteroseismic activities, and take part in the analysis of the large amount of data expected, the KASC will be formed. The purpose of the present mail is to invite you to contribute to this activity, and to gather the information required to organize it in the most efficient way.

The search for extra-solar planets requires exceptional care, to avoid the premature announcement of what might turn out to be a false positive. For this reason the Kepler project, like other similar projects, maintains careful control of access to the data until potential planet detections can be verified. For asteroseismology there is no similar consideration; thus we shall establish an open data policy with the KAI. To make this consistent with the Kepler data policy, the data used for asteroseismology will be pre-processed in such a way that potential information about transits has been removed. This processing will be carried out under the responsibility of Ron Gilliland, before the resulting processed data will be transferred to the Kepler Asteroseismic Science Operations Centre (KASOC) in Aarhus, to be available for the KASC.

If you are interested, I ask you to complete and return the attached form, preferably by e-mail. This includes the nomination of a local contact person for the activity. The form lists several of the activities required in preparation for the mission, and I ask you to indicate your willingness to contribute to these activities, possibly providing further detail in the Comments on how you may contribute. In addition, you are asked to describe briefly the projects that you expect to carry out with the Kepler data. On the basis of the registration we shall provide a letter recognizing your participation which, e.g., can be used in connection with funding proposals. Note that the KASC cannot provide funding for participants, nor will the Kepler project provide funding for these activities. Participants needing funding support will need to seek this from sources available to them through their own local and national agencies.

Given the importance of maintaining careful control over the transit information obtained from Kepler, the Letter of Direction contains a clause requesting a non-disclosure agreement from each individual member of the KASC regarding the unlikely event that transit information will be detected in the asteroseismic data. A form is attached to provide this agreement. I ask you to ensure that this is signed by each person identified as member of the KASC on your form and returned by postal mail, as indicated.

>From the perspective of the Kepler exo-planet programme, the main interest of asteroseismology is the possibility to characterize the central stars of planetary systems. Particularly important is the determination of stellar radii which are required to infer the planet radius from the transit depth. This can be obtained for a substantial fraction of the stars in the exo-planet programme from asteroseismology. Thus we need to develop a reliable and efficient pipeline to provide radius determinations. As member of the KASC you are expected to contribute to this activity.

The Letter of Direction specifies that at least 140 out of the 512 short-cadence targets are to be selected by the KAI for asteroseismic investigation, while in addition the exo-planet program may specify up to 100 planet-hosting targets for asteroseismic characterization. During the initial three months of the mission all 512 short-cadence targets are to be selected by the KAI. The target list for this initial period has to be provided at the latest 4 months before launch. New target lists can be uploaded every three months and must be available 2 - 3 months before the upload. As KASC member you are obviously expected to provide input for the initial and successive target selections.

We plan to hold the first KAI workshop in Paris in October
to discuss the organization of this activity and present
preliminary results of the preparations for the mission. The
KAI is also expected to be discussed at a splinter meeting of
the HELAS II Conference on Helioseismology, Asteroseismology
and MHD Connections in Goettingen, Germany.

Furthermore, we have established the web site
http://astro.phys.au.dk/KASC to serve as a contact to the
KASC.

Below, a preliminary schedule for the pre-launch activities of
KASC is provided.

I look forward to collaborating with you on this exciting
project.

Best regards
Joergen Christensen-Dalsgaard
Head, Danish AsteroSeismology Centre

Preliminary schedule:
May 2007: Invitation to join KASC sent out
June 2007: First version of KASOC pipeline completed
24 August 2007: Splinter meeting at the HELAS II Conference
including a presentation of KAI
29 - 31 October 2007: First KASC Workshop, Paris
January 2008: Review of KASOC analysis procedures
July 2008: Initial target list delivered to the project
November 2008: Launch

# C
# Non-Disclosure Agreement

The following shows the Non-Disclosure Agreement that all initial members of KASC had to sign and return by postal mail.

# Non-disclosure agreement

The undersigned member of the Kepler Asteroseismic Science Consortium (KASC) recognizes that the data made available in the Kepler Asteroseismic Science Operations Centre (KASOC) are intended solely for the use for asteroseismology and that the data have been processed in such a way that all information about planetary transits should have been removed. Should indications suggesting the presence of a transit nonetheless emerge during analysis of the data, with this agreement I undertake not to reveal this information in any form, orally or in writing to anybody, with the exception of Kepler CoI Ron Gilliland (gillil@stsci.edu); Gilliland will then discuss the result with the KeplerPI who will decide on further action.

Name and address:

E-mail:

Date:

Signature

*After completing the form, please return it by postal mail to*

*Hans Kjeldsen*
*Institut for Fysik og Astronomi*
*Aarhus Universitet, bygning 1520*
*Ny Munkegade*
*DK-8000 Aarhus C*
*Denmark*

# D
## Online resources

The following documents, together with an electronic version of this volume, are available via the Zenodo record associated with this booklet (DOI: 10.5281/zenodo.20610469). For each entry, we provide the document name as listed on Zenodo and the date of the corresponding document version.

- Letter of Direction
  (document name: `LoD_signed_orig`; see also § A)
- Non-disclosure agreement (NDA)
  (document name: `Non-Disclosure-KASC`; see also § C)
- KASC Letter of Intent
  (document name: `DASC_KASOC_0012_5`; document date: 20 Nov 2008)
- KAI/KASC Organization and Schedule
  (document name: `DASC_KASOC_0013_6`; document date: 26 Aug 2009)
- KASC target selection procedure: Instructions
  (document name: `DASC_KASOC_0008_5`; document date: 19 Aug 2008)
- Number of targets and target selection procedures
  (document name: `DASC_KASOC_0006_8`; document date: 29 Aug 2008)
- KASC target selection procedure: Instructions for selection of Specific Targets
  (document name: `DASC_KASOC_0023_3`; document date: 15 Dec 2009)
- KASC Data Analysis Procedure and Scientific Publication
  Strategy and Policy
  (document name: `DASC_KASOC_0009_9`; document date: 1 April 2010)
- KASC User Requirements Specifications: Working Group
  Corrected Data
  (document name: `DASC_KASOC_0032_2`; document date: 21 July 2010)
- KASC paper review and submission process
  (document name: `DASC_KASOC_0035_4`; document date: 23 Feb 2011)

- KASC Strategies and Policies in the Extended *Kepler* Mission
  (document name: `DASC_KASOC_0041_7`; document date: 1 Feb 2013)
- Requirement Specifications for TESS: Timing requirements for Asteroseismology
  (document name: `SAC_TESS_0002_6`; document date: 22 July 2014)
- TESS Asteroseismic Science Consortium
  (document name: `SAC_TESS_0003_6`; document date: 16 July 2016)
- TASC Target selection procedure
  (document name: `SAC_TESS_0004_7`; document date: 5 July 2017)
- TASC Publication Strategies and Policies
  (document name: `SAC_TESS_0006_5`; document date: 31 Dec 2017)

# The Kepler and TESS Asteroseismic Consortia

## A historical overview

What is this about? In short: it is about people. About an idea that grew into a consortium, that grew into a community, that grew into a way of doing asteroseismology in the space era. The Kepler Asteroseismic Science Consortium (KASC) and later the TESS Asteroseismic Science Consortium (TASC) are not just organizational structures to access data. They became collaborative ecosystems in which students, postdocs, and senior scientists learned to work together across continents, time zones, and occasionally differing opinions about filters.

*Mikkel N. Lund & Tiago L. Campante*